\documentclass[11pt]{article}

\usepackage{amsmath}
\usepackage{amsfonts}
\usepackage{amssymb}
\usepackage{latexsym}
\usepackage{cite}
\usepackage{lscape} 

\renewcommand{\d}{\mathrm{d}}

\newcommand{\captn}[1]{\vspace{-3ex}\caption{\small #1}}
\newcommand{\ket}[1]{| #1 \rangle}

\DeclareMathSymbol{\mg}{\mathrel}{symbols}{"1D}

\newcommand{\ga}{\alpha}
\newcommand{\gb}{\beta}

\newcommand{\gd}{\delta}

\newcommand{\gf}{\phi}

\newcommand{\gm}{\mu}
\newcommand{\gn}{\nu}

\newcommand{\gl}{\lambda}

\newcommand{\gth}{\theta}
\newcommand{\gvth}{\vartheta}
\newcommand{\gs}{\sigma}
\newcommand{\gt}{\tau}

\newcommand{\gp}{\pi}

\newcommand{\get}{\eta}

\newcommand{\gS}{\Sigma}

\newcommand{\gO}{\Omega}

\newcommand{\cN}{{\cal N}}

\newcommand{\ui}{{\underline i}}

\newcommand{\tv}{{\tilde v}}
\newcommand{\tw}{{\tilde w}}

\newcommand{\tN}{{\tilde N}}

\newcommand{\Tr}{\mbox{Tr}}
\newcommand{\tr}{\text{tr}}

\newcommand{\Id}{\text{\small 1}\hspace{-3.5pt}\text{1}}

\newcommand{\ra}{\rightarrow}

\newcommand{\dsp}{\displaystyle}

\newcommand{\undr}[1]{{\underline{#1}}}

\newcommand{\labl}[1]{\label{#1}}
\newcommand{\Kh}{K\"{a}hler}
\newcommand{\beq}{\begin{equation}}
\newcommand{\eeq}{\end{equation}}
\newcommand{\barr}{\begin{array}}
\newcommand{\earr}{\end{array}}
\newcommand{\equ}[1]{\begin{gather} #1 \end{gather}}
\newcommand{\equa}[1]{\begin{align} #1 \end{align}}

\newcommand{\tabu}[2]{\begin{tabular}{#1} #2 \end{tabular}}
\newcommand{\arry}[2]{\begin{array}{#1} #2 \end{array}}

\newcommand{\pmtrx}[1]{\begin{pmatrix} #1 \end{pmatrix}}

\newcommand{\non}{\nonumber}
\newcounter{oldcounter}

\newcommand{\tga}{{\tilde \alpha}}

\newcommand{\tgf}{{\tilde \phi}}

\newcommand{\tgps}{{\tilde \psi}}
\newcommand{\tget}{{\tilde \eta}}

\newcommand{\Intr}{\mathbb{Z}}
\newcommand{\Cplx}{\mathbb{C}}
\newcommand{\Real}{\mathbb{R}}

\newcommand{\ba}[2]{\[\begin{array}{#2}\label{#1}}
\newcommand{\ea}{\end{array}\]}
\newcommand{\be}{\begin{equation}}
\newcommand{\ee}{\end{equation}}
\newcommand{\bea}{\begin{eqnarray}}
\newcommand{\eea}{\end{eqnarray}}

\newcommand{\E}[1]{\mathrm{E_{#1}}}
\newcommand{\U}[1]{\mathrm{U(#1)}}
\newcommand{\SU}[1]{\mathrm{SU(#1)}}
\newcommand{\SO}[1]{\mathrm{SO(#1)}}
\newcommand{\Sp}[1]{\mathrm{Sp(#1)}}

\newcommand{\brkt}[2]{\bigl[ ^{#1}_{#2} \bigr]}

\newcommand{\rep}[1]{\mathbf{#1}}
\newcommand{\crep}[1]{\overline{\rep{#1}}}
\newcommand{\spin}[2]{\rep{2_{#1}^{{#2}\mbox{-}1}}}
\newcommand{\sm}{{\,\mbox{-}}}

\newcommand{\bibsep}{0ex}

\begin{document}

\thispagestyle{empty}

\begin{flushright}
DESY 04-208\\ 
FTPI-MINN-04/35 \\ 
SNUTP 04-020\\
UMN-TH-2322/04  \\      
hep-th/0410232
\end{flushright}
\vskip 2 cm
\begin{center}
{\Large {\bf 
Heterotic SO(32) model building in four dimensions 
}
}
\\[0pt]
\vspace{1.23cm}
{\large
{\bf Kang-Sin Choi$^{a,}$\footnote{
{{ {\ {\ {\ E-mail: ugha@phya.snu.ac.kr}}}}}}},
{\bf Stefan Groot Nibbelink$^{b,}$\footnote{
{{ {\ {\ {\ E-mail: nibbelin@hep.umn.edu}}}}}}}, 
{\bf Michele Trapletti$^{c,}$\footnote{
{{ {\ {\ {\ E-mail: michele@mail.desy.de}}}}}}},
\bigskip }\\[0pt]
\vspace{0.23cm}
${}^a$ {\it 
School of Physics and Center for Theoretical Physics,
Seoul National University, \\ Seoul 151-747, Korea
 \\}
\vspace{0.23cm}
${}^b$ {\it 
William I. Fine Theoretical Physics Institute, 
School of Physics \& Astronomy, \\
University of Minnesota, 116 Church Street S.E., 
Minneapolis, MN 55455, USA 
\\}
\vspace{0.23cm}
${}^c$ {\it  
Deutsches Elektronen Synchrotron (DESY), \\
Notkestrasse 85, D-22607 Hamburg, Germany
}
\bigskip
\vspace{1.4cm} 
\end{center}
\subsection*{\centering Abstract}

Four dimensional heterotic $\SO{32}$ orbifold models are classified
systematically with model building applications in mind. 
We obtain all $\Intr_{3}$, $\Intr_7$ and $\Intr_{2N}$  
models based on vectorial gauge shifts. The resulting gauge groups are
reminiscent of those of type--I model building, as they always take the
form 
$\SO{2n_0} \times \U{n_1} \times \ldots \times \U{n_{N\sm 1}}
\times \SO{2n_N}$.
The complete twisted spectrum is determined simultaneously for all
orbifold models in a parametric way depending on 
${\rm n_0},\ldots,{\rm n_N}$, rather than on a model by model
basis. This reveals  interesting patterns in the twisted states: They
are always built out of vectors and anti--symmetric tensors of the
$\U{n}$ groups, and either vectors or spinors of the $\SO{2n}$
groups. Our results may shed additional light on the $S$--duality
between heterotic and type--I strings in four dimensions. As a
spin--off we obtain an $\SO{10}$ GUT model with four generations from
the $\Intr_4$ orbifold.

\newpage

\setcounter{page}{1}

\section{Introduction and summary}
\labl{sc:intro}

Since the mid eighties there have been many studies to the physics of extra 
dimensions. This route was first considered seriously with the
development of superstrings, and in particular when it was realized that
the heterotic $\E{8}\times\E{8}$ string \cite{Gross:1985fr,Gross:1985rr}
can give rise to four dimensional phenomenology \cite{Candelas:1985en}
by considering Calabi--Yau or orbifold compactification
\cite{dixon_85,Dixon:1986jc}. Conformal field theories on the latter
spaces  \cite{Dixon:1987qv}  are particularly simple since they are free. 
The extension to orbifolds with gauge field background, or Wilson
lines, has been first investigated in ref.\ \cite{Ibanez:1987pj}. 
A major part of the literature has been devoted to the heterotic
$\E{8}\times\E{8}$ theory, that was the first string theory to be
considered seriously for Standard Model (SM) phenomenology. This was
due mainly to the fact that even for the simplest standard embedding
of the spin connection in the gauge group, Grand Unified Theory (GUT)
gauge groups arise from one of the $E_8$ gauge groups, while the other
$\E{8}$ group can be considered as part of a hidden sector that might
be responsible for supersymmetry breaking by gaugino condensation
\cite{Ferrara:1982qs,Derendinger:1985kk,Dine:1985rz}. After these 
developments there have been many efforts to obtain a full picture of all 
possible gauge groups that can arise from heterotic $\E{8}\times\E{8}$ orbifolds 
\cite{Casas:1989wu,Katsuki:1989kd,Katsuki:1990bf,Kawamura:1996zu,Choi:2003pq,Choi:2004vb}. 
For recent investigations to heterotic string Supersymmetric Standard
Model (MSSM) and GUT model building we
refer to \cite{Giedt:2000bi,Giedt:2001zw,Forste:2004ie,Nilles:2004ej}
and references therein.

The study of string phenomenology turned a different direction with the
construction of $D$--branes \cite{Pradisi:1988xd,Polchinski:1995mt} in
type--II string 
theory. A stack of $D$--branes gives rise to $\U{n}$ or $\SO{2n}$
gauge groups, and therefore models with various stacks of branes lead
to effective theories with products of such gauge groups. The
cancellation of $RR$--flux tadpoles selects consistent $D$--brane models
\cite{Gimon:1996rq}. In particular, the type--I string can be viewed as
a type--II orientifold. By considering branes at
angles \cite{Blumenhagen:1999ev} it has been possible to construct
orientifold models with similar gauge group and spectrum as the
Standard Model or its supersymmetric extension 
\cite{Forste:2000hx,Forste:2001gb,Blumenhagen:2001te,Cvetic:2001nr}. 
An interesting aspect is that type--I string and the heterotic
$\SO{32}$ string are related via a strong/weak duality in ten dimensions
\cite{Polchinski:1995df}. Therefore also the phenomenology of the
heterotic $\SO{32}$ orbifold models should be studied in detail. 
Some steps in that direction have been taken refs.\
\cite{Giedt:2003an,Giedt:2004wd}, where $\Intr_3$ orbifolds with
Wilson lines were considered. And \cite{Dine:2004dk} investigates
discrete symmetries like CPT in the ($\SO{32}$) string context. 
We seek to obtain a classification
of more general orbifolds in the heterotic $\SO{32}$ string context,
but for the sake of simplicity we ignore the possibility of Wilson lines. 
This investigation may shed additional light on the S--duality between
heterotic and type--I strings in four dimensions.

Another motivation for our pursuit of a classification of heterotic
$\SO{32}$ models is that they may be useful extension of field theory
models of extra dimensions. In recent years there has been a lot of
interests in five, six and higher dimensional orbifold field theories 
\cite{Arkani-Hamed:1998rs,Antoniadis:1998ig,Arkani-Hamed:1998nn,
Barbieri:2000vh,Delgado:1999sv,Delgado:1998qr}
and orbifold GUTs \cite{Hebecker:2001jb,Hall:2001xr,
Asaka:2001eh,Hebecker:2003jt}, making use of
split multiplets for the Higgs \cite{Kawamura:2000ev}.
Essentially all these models are non--renormalizable and therefore
require some form of ultra--violet completion. At the moment the only
candidates for complete theories in extra dimensions come from string
theory. A concrete example for some orbifold GUTs have been obtained 
from heterotic string theory, by taking a--symmetric limits of some of
the radii of the $\Intr_6$ orbifold \cite{Kobayashi:2004ud,Kobayashi:2004ya},
for a general investigation to the scales in such a scenario see also 
\cite{Hebecker:2004**}.
Many of the heterotic $\SO{32}$ models, that we classify in this work,
may be used for field theoretical investigations in a similar way.

The main results of our work can be summarized as: 
We give a systematic classification of four dimensional heterotic
$\SO{32}$ orbifold models. We obtain all $\Intr_3$, $\Intr_7$ and
$\Intr_{2N}$ models based on vectorial gauge shifts. The resulting
gauge groups are reminiscent of those obtained in type--I model building as
they generically take the form: 
\(
\SO{2n_0} \times \U{n_1} \times \ldots \times \U{n_{N\sm 1}}\times
\SO{2n_N}. 
\) 
Most classification works for $\E{8}\times\E{8}$ orbifolds stop here,
once the resulting gauge group has been obtained. We continue to
determine the complete twisted spectrum simultaneously for all
orbifold models, rather than on a model by model basis. This reveals
interesting patterns in the twisted states that are manifestly
portrayed in our classification tables. For example we give explicit
mappings between various twisted spectra, which greatly reduces the
classification effort. The paper is outlined as follows:

In section \ref{sc:ModZ3} and \ref{sc:ModZ4} we review the
classification of $\Intr_3$ and $\Intr_4$ 
orbifold models respectively, as simple examples of odd and even order
orbifold models and to fix precisely the notation. We also compute the
anomaly polynomial in order to have a consistency check
on the spectrum. The anomaly analysis allows us to understand how the
only heterotic model having a type--I counterpart in the $\Intr_3$
case is free from irreducible anomalies already at the level of
untwisted states, while the other models, with one excepction, do not
have this feature. Interestingly, the exception has gauge group equal
to the original $SO(32)$, which, to our knowledge, has no type--I
dual.

Section \ref{sc:ModZ2N}, the core of our paper, explains the
details of our classification procedure. We first reduce the problem
of finding modular invariant shifts to an exercise in linear
algebra. In order to perform the classification, we restrict ourselves to
those twisted states, that cannot be obtained from other twisted
states by orbifolding. By exploiting spectral flow we bring all weights in a
standard form so that classification becomes very simple. In this way
we show that the structure of the twisted states is always the
same. We recompute the twisted $\Intr_3$ and $\Intr_4$ to illustrate
how efficient our classification procedure is. We complete the
classification prescription by returning to the reducible twisted
states and explain how to compute them from the irreducible ones using
primarily field theory methods.

In section 5 we give various possible applications and extensions of
our results. We first explain classification procedure can be easily
extended to $\Intr_N\times\Intr_{N'}$ orbifold models and to
$E_8\times E_8$ heterotic string. Next, we consider the 
heterotic/type--I $S$--duality in the light of our classification on
the heterotic side. For each of the odd order orbifolds ($\Intr_3$
and $\Intr_7$) there are only two models free of irreducible anomalies at
the untwisted spectrum level. But only one has a type--I counterpart,
while the other, with gauge group $SO(32)$, does not have a type--I
dual.

In appendix A we describe  the modular invariant partition function
of a generic $\Intr_N$ orbifold model that is basis of our
classification. In the other appendices we give
the classification of $\Intr_6$, $\Intr_7$ and $\Intr_8$ models. There
one finds a classification of the modular invariant shifts and tables
of the irreducible spectra.  For $\Intr_7$ model we also give the
anomaly polynomial corresponding to the irreducible anomaly.

\subsection*{Acknowledgments}

We would like to thank E.\ Dudas, A.\ Hebecker, T.\ Gherghetta, 
M.\ Laidlaw, O.\ Lebedev, M.\ Ratz, and G.\ Shiu  for stimulating
discussions, and H.P.\ Nilles for communications on related topics.
SGN would like to thank the Theory division of Seoul  National
University for their kind hospitality during part of this project. MT
would like to thank the William I.\ Fine Theoretical Physics Institute
at the University of Minnesota for their kind hospitality during part
of this project. 

We would like to thank the authors of \cite{Nilles:2006np}, H.P.\
Nilles, S.\ Ramos--S'anchez, P.\ Vandrevange and A.\ Wingerter,  for
pointing out that our shift vectors do not constitute the most general
form and made us aware of a few typos in table \ref{tb:SpecZ4} of a
previous version of this paper.

The work of SGN is supported in part by the Department of Energy under
contract DE--FG02--94ER40823 at  the University of Minnesota. 
KSC is supported in part by the 
KOSEF Sundo Grant R02-2004-000-10149-0,
KOSEF ABRL Grant to the Particle Theory Group of SNU.

\subsection{Technical preliminaries}
\labl{sc:tech}

We consider the heterotic $\SO{32}$ theory on a six dimensional torus
$T^6$.  (We will ignore the possibility of having various radii for
these tori for simplicity.) To obtain a chiral spectrum this torus must
be orbifolded. To this end we write $T^6 = T^2 \times T^2 \times T^2$ 
parameterized by the complex coordinates $z^i$, $i = 1,2,3$. A 
compact orbifold is obtained by the identification 
\equ{
z^i \ra e^{2\pi i \, \gf_i} z^i,
} 
due to a discrete Abelian group that we take to be $\Intr_N$. 
Crystallography of the torus lattices only allow $N=2,3,4,6,7,8$ and
$12$. Preservation of supersymmetry leads to the requirement that 
\equ{
\sum_i \gf_i \equiv 0, 
}
where the equivalence relation $a \equiv b$ indicates that $a$ and $b$
differ by an integer. All the standard shifts are chosen such that
equality holds
\equ{
\arry{c}{
\phi_{\Intr_{3}}=\frac 13 (1,1,\mbox{-}2),\qquad 
\phi_{\Intr_{4}}=\frac 14 (1,1,\mbox{-}2),\qquad 
\phi_{\Intr_{6}\mbox{\scriptsize--I}}=\frac 16 (1,1,\mbox{-}2),
\\[2ex]
\phi_{\Intr_{7}}=\frac 17 (1,2,\mbox{-}3),\qquad 
\phi_{\Intr_{8}\mbox{\scriptsize--I}}=\frac 18 (1,2,\mbox{-}3).
}
\labl{StSpaceShift} 
}

In string the spacetime coordinate become bosonic fields $X^M$ on the
string worldsheet. As the heterotic theory only contains closed
strings, the ten coordinate field $X^M$ can be expanded 
\equ{
X^M(\gs_0,\gs_1) = \sum_{k} 
\Big( e^{2\pi i\, k(\gs_1-\gs_0)} \ga^M_{k} + e^{2\pi i\, k(\gs_1+\gs_0)} \tga^M_{k}
\Big)
}
in both left-- and right--moving oscillators, $\ga^M_k$ and $\tga^M_k$,
respectively. Excited states are created from the vacuum $\ket{0}$ by
acting on it with these oscillators. To facilitate the identification with field
theoretical compactification it is useful to define such states with
lower indices. For example,
\equ{
\ket{_M} = \get_{MN} \ga_{-k}^N \ket{0}, 
}
gives a target space vector. This is compatible with the fact that gauge
fields being connections have their spacetime index downstairs:
$A_M$. On the right--moving side the worldsheet theory is
supersymmetric, and contains ten real fermions $\tgps^M$. Instead, on
the left--moving side the theory can be described as having 16 complex
fermions $\gl^I$ and hence this side is not supersymmetric. Here $I$
labels 16 Cartan generators of $\SO{32}$. 
The orbifold action is embedded in the $\gl^I$ through a 16 dimensional vector
$v=(v^I)$
\equ{
\gl^I \ra e^{2\pi i \, v_I} \gl^I.
} 
This complete our technical introduction.

\section{$\boldsymbol{\Intr_3}$ models}
\labl{sc:ModZ3}

We investigate the heterotic $\SO{32}$ theory compactified on the six
dimensional orbifold $T^6/\Intr_3$ with space action $\phi=\frac 13 (1,1,\mbox{-}2)$. 
The orbifold action on the gauge group is defined by the shift vector 
\equ{
v =  \mbox{$\frac 13$} \big( 0^{16-3n},1^{2n}, 
\mbox{-}2^{n} \big), 
\qquad 
0 \leq n \leq 5, 
\labl{shift3}
}
satisfying the modular invariance constraint 
\(
\frac 32 v^2 \equiv 0.
\)
For $n>1$ this is a multiple
embedding of the so--called standard embedding with $n=1$. 
(In the $\E{8} \times \E{8}$ such multiple embeddings can always be 
reduced to simpler embeddings. This is not the case in the $\SO{32}$
theory because, contrary to the $\E{8}$ case, there are no spinorial
weights in which one can perform Weyl reflections.) 
At the fixed points the $\SO{32}$ gauge symmetry is broken to 
\equ{
\SO{32} \ra \SO{32 - 6n} \times \U{3n}. 
}
The untwisted matter can determined by considering the branching of
the adjoint of $\SO{32}$: 
\equ{
\rep{ 496 } \ra 
(\rep{ \mbox{$\frac 12$} (32 \!-\! 6n) (31 \!-\! 6n) }, \rep{1})_0 
+  (\rep{1}, \rep{ (3n)^2 })_0 
+ (\rep{1}, \rep{ \mbox{$\frac 12$} 3n (3n \!-\! 1) } )_{2}
+ (\rep{1}, \crep{ \mbox{$\frac 12$} 3n (3n \!-\! 1) } )_{\mbox{-}2}
\non \\[2ex] 
+ ( \rep{32\!-\! 6n}, \rep{3n} )_{1} 
+ ( \rep{32\!-\! 6n}, \crep{3n} )_{\mbox{-}1}.  
}
The $\U{1}$ charge operator $q =( 0^{16-3n}, 1^{3n} )$ is fixed by the
requirement that the $\SO{32-6n}$ roots 
\( 
( \underline{\pm 1^2, 0^{14-3n}}, 0^{3n} )
\)
and the $\SU{3n}$ roots 
\(
( 0^{16-3n}, \underline{1,\mbox{-}1, 0^{3n -2}} )
\) 
are neutral. The underline denotes all possible permutations. 
In the case $n=5$ there are no $\rm{SO}$ roots but
instead there is an additional charge operator 
$q' = ( 1, 0^{15} )$. Of course for $n=0$ the gauge group is $\SO{32}$.
The $\SO{32}$ roots representation of the untwisted matter reads 
\equ{
v \cdot w - \frac 13 \equiv 0~:~
\left\{ 
\arry{rl}{
( \rep{32\!-\! 6n}, \rep{3n} )_{1} ~: &  
w= ( \underline{\pm 1, 0^{15-3n}}, \underline{1, 0^{3n-1}} ),
\\[2ex]
 (\rep{1}, \crep{ \mbox{$\frac 12$} 3n (3n\!-\! 1) } )_{\mbox{-}2} ~: & 
w = ( 0^{16-3n}, \underline{\mbox{-}1^2, 0^{3n-2}} ).
}
\right. 
}
These untwisted states
form a triplet under the $\SU{3}_H$ holonomy group; this leads to
a multiplicity of three for the untwisted spectrum. 
This untwisted matter results in the anomaly polynomial 
\equ{
I_{6|u} = - \frac 16 27 (4 - n ) \, \tr\, F_{\rm{SU}}^3 
+ \frac 16 27 \!\cdot\! 6n \!\cdot\! (n -2) \, F^3 
- \frac 1{48} 27 n ( 3n -11) \, F\,  \tr\, R^2 
\non \\[2ex] 
+ \frac 12 36 (n-3) \, F\,  \tr\, F_{\rm{SU}}^2 
- \frac 12 9 n \, F\,  \tr\, F_{\rm{SO}}^2,
\labl{unAnom}
}
that encodes the structure of the pure and mixed anomalies. 
(We recall that for an anti--symmetric tensor representation
$[\rep{r}]_2$ of any representation $\rep{r}$ of dimension $|\rep{r}|$
the trace identities hold: 
$\tr_{\rep{r}_2} F^2 = ( |r| -2 ) \tr_\rep{r} F^2$ and 
$\tr_{\rep{r}_2} F^3 = ( |r| -4 ) \tr_\rep{r} F^3$.)
Here $F_{\rm{SO}}, F_{\rm{SU}}$ and $F$ are the field strength
two--forms  of the $\SO{32-6n}, \SU{3n}$ and $\U{1}$ gauge 
symmetries, respectively, and $R$ denotes the curvature two--form.  
The traces of the $\SO{32-6n}$ and $\SU{3n}$ field strengths are 
evaluated in the vector representation unless otherwise indicated. 
In the $n=5$ model also the field strength $F'$ is present in the
anomaly polynomial. As the pure $\U{1}'$ anomaly and all mixed
anomalies involving a single $F'$ vanish immediately, it only gives 
a single extra term $-45 \, F\, {F'}^2$ replacing the contribution 
$-\frac 12 9n\, F\, \tr\, F_{\rm{SO}}^2$.

\begin{table}
\[
\arry{| c | cc | cc | cc |}{
\hline &&&&&& \\[-2ex]  
n & \rep{1}_H & \text{weights} & \crep{3}_H &  \text{weights} 
& \rep{3}_H+\crep{6}_H & \text{weights}
\\[1ex] \hline &&&&&& \\[-2ex] 
0 && && &\rep{1} & (0^{16}) 
\\[1ex] \hline &&&&&& \\[-2ex] 
1 & (\rep{1}, \rep{1}),~ (\rep{26}, \rep{1}) 
& (0^{13}, \mbox{-}1^3 ),~ (\underline{\pm 1, 0^{12}}, 0^3 ) 
& (\rep{1}, \crep{3}) & (0^{13}, \underline{ \mbox{-}1, 0^2} ) 
&& \\[1ex] \hline &&&&&& \\[-2ex] 
2 & (1, \crep{15}) & (0^{10}, \underline{\mbox{-}1^2, 0^4}) 
& (\rep{1}, \rep{1}) & (0^{16}) 
&& \\[1ex] \hline &&&&&& \\[-2ex] 
3 & ( \rep{1}, \crep{9} ) & (0^7, \underline{\mbox{-}1, 0^8} ) 
&& 
&&\\[1ex] \hline &&&&&& \\[-2ex] 
4 &  (\rep{1}, \rep{1}),~ (\rep{8}_+, \rep{1}) & 
(0^{16}),~ 
(\underline{ \pm \mbox{$\frac 12$}^{4}}, \mbox{-$\frac 12$}^{12})
&&
&&\\[1ex] \hline &&&&&& \\[-2ex] 
5 & \rep{15} & 
(\mbox{$\frac 12$}, 
\underline{ \mbox{-$\frac 12$}^{14}, \mbox{$\frac 12$}}) 
& \rep{1} & (\mbox{-$\frac 12$}^{16})
&&\\[1ex] \hline 
}
\]
\captn{The twisted states and their weights are summarized for the
five $\SO{32-6n}\times \SU{3n}$ models, dropping the multiplicity of
27 due to the fixed points. 
\labl{tb:TwStates}}
\end{table}

The twisted states either are singlets or triplets of the $\SU{3}_H$
holonomy group. They are determined by the relations 
\equ{
\rep{1}_H~:~~ \frac 12 ( w + \tv )^2 = \mbox{$\frac 23$}, 
\quad 
\crep{3}_H~:~~   \frac 12 ( w + \tv )^2 = \mbox{$\frac 13$},
\quad 
\rep{3}_H + \crep{6}_H~:~~   \frac 12 ( w + \tv )^2 = 0,
} 
where $\tv = ( 0^{16-3n}, \mbox{$\frac 13$}^{3n} )$. 
Here the $\crep{6}_H$ representation arises as the symmetric
two--index tensor  representation. This and the representation
$\rep{3}_H$ can only appear in the full $\SO{32}$ orbifold, i.e.\ the
case $n=0$.

 In table 
\ref{tb:TwStates} we have collected the representations of the twisted
states and indicated their weights.  By exploiting spectral flow, 
we have chosen to use the vector $\tv$ instead of the original shift
vector $v$ to obtain 
single forms for the weights in this table. The $\U{1}$ charges of the
twisted states are computed as $q\cdot (w + \tv)$. (If one use the
original $v$ the corresponding weights will be different but the
resulting charge is the same.) The anomaly polynomial due to 
the twisted states  in the representation $(\rep{r}, \rep{s})_q$ reads  
\equ{
I_{6|(\rep{r}, \rep{s})_q} = 27 \Big[ 
- \frac 16 |\rep{r}| \, \tr_{\rep{s}} F_{\rm{SU}}^3 
- \frac 16 |\rep{r}| |\rep{s}| q^3 \, F^3 
+ \frac 1{48} |\rep{r}| |\rep{s}| q \, F \, \tr\, R^2  
\non \\[2ex] 
- \frac 12  |\rep{r}| q \, \tr_{\rep{s}} F_{\rm{SU}}^2 
- \frac 12  |\rep{s}| q \, \tr_{\rep{r}} F_{\rm{SO}}^2 \Big].  
}
For the holonomy triplets there is an extra multiplicity 
factor of three. 

In order that the anomalies can be canceled by the Green--Schwarz
mechanism, it is necessary that the total anomaly polynomial $I_6$
factorizes as 
\equ{
I_6 = c\, F X_{4|4D} = c\,  F \Big[ 
\tr\, R^2 - \tr\, F_{\rm{SO}}^2 - 2 \tr\, F_{\rm{SU}}^2 -  6n\, F^2 
\Big]. 
\labl{fac4Dhet}
}
The relative coefficients of the traces are fixed because the four
dimensional Green--Schwarz mechanism is remnant of this 
mechanism in ten dimensions, where the field strength $H$ of 
the anti--symmetric tensor $B$ fulfills the anomalous Bianchi 
identity
\( 
\d H = X_4 = \tr\, R^2 - \tr\, F_{\SO{32}}^2. 
\) 
At the four dimensional fixed points this four form is restricted 
to $X_{4|4D}$ given in \eqref{fac4Dhet}. The presence of the factor of
2 in front of the $\SU{3n}$ trace is obtained by taking to account the 
indices of $\rm{SO}$ and $\rm{SU}$ groups. This in particular fixes
the coefficient in front of the $F^2$ term: It is given as the trace
of the $\SO{2}$ generator identified by $q$ which using the $\U{1}$ --
rather than the $\SO{2}$ -- normalization gives the factor $2 \cdot 3n$. The
coefficients $c$ are tabulated in table \ref{tb:FacCoeff}. In the $n=5$
model there  are two $\U{1}$'s, so that the factorization
takes the form 
\equ{
I_{6|n=5} = (c\,  F + c'\, F')  \Big[ 
\tr\, R^2 - 2 \tr\, F_{\rm{SU}}^2 - 6n\, F^2 - 2'\, {F'}^2
\Big]. 
\labl{fac4Dhet5}
}
The appropriate coefficients $(c, c')$ are also displayed in table 
\ref{tb:FacCoeff}. Observe that from this factorization it follows
that of the linear combinations of $\U{1}$ generators 
\equ{
q_{a} = 225 q - 27 q', 
\qquad 
q_n = 27 q' + 225 q, 
}
only $q_a$ is anomalous.

\begin{table} 
\begin{center} \begin{tabular}{| c | c | c | c |}
\hline &&& \\[-2ex]  
$n$ & gauge group & untwisted (x 3) & twisted (x 27)
\\[1ex] \hline &&& \\[-2ex] 
0 & $\SO{32}$ &  & $9 (\rep{1})$  
\\[1ex] \hline &&& \\[-2ex] 
1 & $\SO{26}\times \SU{3} \times \U{1}$
& ${\bf (26,3)_{1}}+{\bf (1,3)}_{\mbox{-}2}$
& $3{\bf (1,\bar 3)_0}+ {\bf(1,1)}_{\mbox{-}2}+ {\bf (26,1)}_{1}$
\\[1ex] \hline &&& \\[-2ex] 
2 & $\SO{20}\times \SU{6} \times \U{1}$
 & $ {\bf (20,6)}_{1} + {\bf (1,\overline{15})}_{\mbox{-}2} $
& $ 3{\bf (1,1)}_{2} + {\bf (1,\overline{15})}_{0}$
\\[1ex] \hline &&& \\[-2ex] 
3 & $\SO{14}\times \SU{9} \times \U{1}$
& ${\bf (14,9)}_{1} + {\bf (1,\overline{36})}_{\mbox{-}2}$
& ${\bf (1,\overline{9})}_{2}$
\\[1ex] \hline &&& \\[-2ex] 
4 & $\SO{8} \times \SU{12} \times \U{1}$
& ${\bf (8,12)}_{1} + {\bf (1,\overline{66})}_{\mbox{-}2}$
& ${\bf (1,1)_{4}}+ {\bf (8_+,1)}_{\mbox{-}2} $
\\[1ex] \hline &&& \\[-2ex] 
5 & $\SU{15} \times \U{1} \times \U{1}'$
& $ ({\bf  15})_{1,\mbox{-}1}+ ({\bf 15})_{1,1} 
+  (\overline{\bf 105})_{\mbox{-}2, 0} $
& $ 3 {\bf (1)_{\mbox{-}\frac 52, \mbox{-}\frac 12}} 
+ ({\bf 15})_{\mbox{-}\frac 32,\frac 12} $
\\[1ex] \hline 
\end{tabular} \end{center}
\captn{
This table gives the complete spectrum of the $\SO{32}$ heterotic
$\Intr_3$ orbifold models with the gauge shift vector 
\(
v =  \mbox{$\frac 13$} \big( 0^{16-3n},1^{2n}, 
\mbox{-}2^{n} \big)
\).
\labl{tb:fullSpec}}
\end{table}

\begin{table}
\[
\arry{| c | c | c | c | c | c |}{ 
\hline &&&&& \\[-2ex]  
n & 1 & 2 & 3 & 4 & 5 
 \\[1ex] \hline &&&&& \\[-2ex] 
c &18 & 9 & \mbox{$\frac{27}2$} & \mbox{-}9 & 
( \mbox{-$\frac{225}8$}, \mbox{$\frac{27}8$} ) 
\\[1ex] \hline 
}
\]
\captn{The factorization coefficients defined in \eqref{fac4Dhet}
are tabulated for the five $\Intr_3$ orbifold models. As the theory
$n=5$ contains two $\U{1}$'s there in total four coefficients; the
factorization in that case is given in \eqref{fac4Dhet5}. 
\labl{tb:FacCoeff}}
\end{table}


\section{$\boldsymbol{\Intr_4}$ models}
\labl{sc:ModZ4}

\newcommand{\Zfour}[1]{
\begin{landscape}
\begin{table} 
\begin{center} 
\scalebox{#1}{\mbox{
\begin{tabular}{| cc | c | c | c | c |}
\hline &&&&& \\[-2ex]  
$ n_1$ & & 6D gauge group & 6D untwisted $\rep{R}$  
& 6D twisted $\rep{D}, \rep{S}$  & 
\\[1ex] \hline  &&&&& \\[-2ex] 
& $n_2$ & 4D gauge group & 4D untwisted $\rep{R}_{i=1,2}$  & 
6D twisted on $T^2/\Intr_2$ & 4D twisted $\rep{T}$ 
\\[1ex] \hline\hline  &&&&& \\[-2ex] 
$2$ &  & $\SO{28} \times \SO{4}$ & 
$(\rep{28},\rep{4}) + 4(\rep{1},\rep{1})$ & 
$20 (\rep{1},\rep{2_-}) + 5 (\rep{28},\rep{2_+})$ 
& 
\\[1ex] \hline  &&&&& 
\\[-2ex]&&&&& $5 (\rep{1},\rep{1})_{\frac 12,\frac 12} 
+ 2 (\rep{1},\crep{2})_{\sm\frac12,\sm\frac 12} $ 
\\[-2ex] 
& $1$  & $\SO{26} \times \U{2} \times  \U{1}'$ &
$(\rep{26}, \rep{2})_{1,0} + (\rep{1}, \crep{2})_{\sm 1,\pm 1}$
& $2(\rep{1}, \rep{2})_{0,0} + (\rep{26},\rep{1})_{1,0}+ (\rep{1},\rep{1})_{-1,\pm 1}$ 
\\[-1ex] &&&&& 
$+ (\rep{1},\rep{1})_{\sm\frac 32,\frac 12}
+ (\rep{26},\rep{1})_{\frac 12,\sm \frac 12}$
\\[.5ex] 
&&&&& \\[-2ex] 
&$3$  & $\SO{22} \times \U{2} \times \SO{6}'$ &
$(\rep{22},\rep{2},\rep{1})_{1} + (\rep{1},\crep{2},\rep{6})_{\sm 1}$
& $2 (\rep{1}, \rep{2}, \rep{1})_0 + (\rep{22},\rep{1},\rep{1})_{\sm 1} 
        + (\rep{1},\rep{1},\rep{6})_1 $ 
& $2 (\rep{1}, \rep{1}, \rep{4}_+)_{\frac 12} 
+ (1, \rep{2}, \rep{4}_-)_{\sm \frac 12}$
\\[1ex] 
&&&&& \\[-2ex] 
&$5$ & $\SO{18} \times \U{2} \times \SO{10}' $ &
$(\rep{18},\rep{2},\rep{1})_{1} +
(\rep{1},\crep{2},\rep{10})_{\sm 1}$
&  $2 (\rep{1}, \rep{2}, \rep{1})_0 + (\rep{18}, \rep{1},\rep{1})_{1}
        +(\rep{1}, \rep{1},\rep{10})_{\sm 1} $ 
& $(\rep{1},\rep{1},\rep{16}_+)_{\frac 12}$ 
\\[1ex] 
&&&&& \\[-2ex] 
&$7$ & $\SO{14} \times \U{2} \times \SO{14}'$ &
$(\rep{14},\rep{2},\rep{1})_{1} + (\rep{1},\crep{2},\rep{14})_{\sm 1}$
&  $2 (\rep{1}, \rep{2}, \rep{1})_0 + (\rep{14},\rep{1},\rep{1})_{\sm 1}
        +(\rep{1},\rep{1},\rep{14})_{ 1} $
& 
\\[1ex] \hline\hline &&&&& \\[-2ex] 
6 & & $\SO{20} \times \SO{12}$ & 
$(\rep{20},\rep{12})+4 (\rep{1},\rep{1})$ & 
$5 (\rep{1},\rep{32}_+)$ 
& 
\\[1ex] \hline &&&&& \\[-2ex] 
&$0$ & $\SO{20} \times \U{6}$ &
$(\rep{20},\rep{6})_{1}$
& $(\rep{1},\rep{1})_{\sm 3} + (\rep{1}, \crep{15})_{1}$ 
& $5 (\rep{1}, \rep{1})_{\frac 32} + (\rep{1}, \crep{15})_{\sm \frac 12}$ 
\\[1ex] 
&&&&& \\[-2ex] 
&$2$ & $\SO{16} \times \U{6}  \times \SO{4}'$ &
$(\rep{16},\rep{6},\rep{1})_{1} + (\rep{1},\crep{6},\rep{4})_{\sm 1}$ 
& $(\rep{1},\rep{1},\rep{1})_{3} + (\rep{1}, \rep{15},\rep{1})_{\sm 1}$ 
& $2(\rep{1}, \rep{1}, \rep{2}_+)_{\frac 32} 
+ (\rep{1}, \crep{6}, \rep{2}_-)_{\frac 12}$ 
\\[1ex] 
&&&&& \\[-2ex] 
&$4$ & $\SO{12} \times \U{6}  \times \SO{8}'$ &
$(\rep{12},\rep{6},\rep{1})_{1} + (\rep{1},\crep{6},\rep{8}_v)_{\sm 1}$ 
& $(\rep{1},\rep{1},\rep{1})_{\sm 3} + (\rep{1}, \crep{15},\rep{1})_{1}$ 
& $(\rep{1}, \rep{1}, \rep{8}_+)_{\frac 32}$
\\[1ex] \hline\hline &&&&& \\[-2ex] 
10 & & $\SO{12} \times \SO{20}$ & 
$(\rep{12},\rep{20}) + 4(\rep{1},\rep{1})$ & 
$5 (\rep{32}_-,\rep{1})$ 
&
\\[1ex] \hline &&&&& \\[-2ex] 
&$1$ & $\SO{10} \times \U{10} \times \U{1}'$ &
$(\rep{10}, \rep{10})_{1,0} + (\rep{1}, \crep{10})_{\sm 1,\pm 1}$
& $(\rep{16}_-,\rep{1})_{0,\sm \frac 12}$ 
& $2(\rep{1},\rep{1})_{\frac 52,\frac 12}
+ (\rep{1}, \crep{10})_{\frac 32,\sm \frac 12} $
\\[1ex] 
&&&&& \\[-2ex] 
&$3$ & $\SO{6} \times \U{10} \times \SO{6}'$ &
$(\rep{6},\rep{10},\rep{1})_{1} + (\rep{1},\crep{10},\rep{6})_{\sm 1}$
& $(\rep{4}_-, \rep{1}, \rep{4}_+)_0$ 
& $(\rep{1},\rep{1}, \rep{4}_+)_{\frac 52} 
+ (\rep{4}_-,\rep{1}, \rep{1})_{\sm \frac 52}$
\\[1ex] \hline\hline  &&&&& \\[-2ex] 
$14$ &  & $\SO{4} \times \SO{28}$ & $(\rep{4},\rep{28})$ & 
$20 (\rep{2_+},\rep{1}) + 5(\rep{2_-},\rep{28})$ 
& 
\\[1ex] \hline &&&&& \\[-2ex] 
&$0$  & $\SO{4} \times \U{14}$ &
$(\rep{4},\rep{14})_{1}$ 
& $ 2(\rep{2_+},\rep{1})_0 + (\rep{2}_-,\crep{14})_{-1}$ 
& $2 (\rep{1}, \rep{1})_{\frac 72} + (\rep{2}_+,\rep{1})_{\sm \frac 72} $
\\[1ex] \hline 
\end{tabular} 
}}
\end{center}
\captn{
The six and four dimensional gauge groups are tabulated  of the
$\SO{32}$ heterotic $\Intr_4$ orbifold models defined by the gauge
shift vector 
\(
v =  \mbox{$\frac 14$} 
\big( 0^{n_0}, 1^{n_1}, 2^{n_2} \big), 
\)
$n_0 = 16-n_1-n_2$. The six
dimensional (half) hyper multiplets included the multiplicity factors
that count the number of independent $T^4/\Intr_2$ fixed points within
$T^6/\Intr_4$. The four dimensional twisted states and zero modes of
the twisted states on $T^2/\Intr_2$ complete the table. This
table does not give the complete four dimensional spectrum, only the
chiral part, relevant for anomaly considerations. 
\labl{tb:SpecZ4}}
\end{table} 
\end{landscape} 
}

Next we study the heterotic $\SO{32}$ theory on $T^6/\Intr_4$. The
analysis is to a large extent similar to the $\Intr_3$ case except
that, as $T^6/\Intr_4$ is an even order orbifold, it has six
dimensional hyper surfaces at the fixed points of the orbifold
$T^4/\Intr_2$. Four of them are orbifolds $T^2/\Intr_2$ and the other
$12$ are $T^2$'s that are mapped pairwise to each other. The fixed
points of the four $T^2/\Intr_2$ combined coincides with the fixed
points of the original $T^6/\Intr_4$. The situation is very similar to the
heterotic $\E{8}$ theory on the same orbifold studied in ref.\ 
\cite{Nibbelink:2003rc}. The spacetime and gauge shift vectors are 
generically given by 
\equ{ 
\gf = \mbox{$\frac 14$} \big( 1^2, \mbox{-}2 \big), 
\quad 
v =  \mbox{$\frac 14$} 
\big( 0^{n_0}, 1^{n_1}, 2^{n_2} \big), 
}
with $n_0 = 16-n_1-n_2$. (We ignore the possibility of having
spinorial shifts as well as some exceptional cases mentioned after
\eqref{shift2Nvec}.)  For the shift vector $v$ the resulting six and
four dimensional gauge groups are   
\equ{
\SO{32} 
\ra \SO{2(n_0 + n_2)}  \times \SO{2n_1}
\ra \SO{2n_0} \times \U{n_1} \times \SO{2n_2}'. 
\labl{Gauge4}
}
Notice that by taking $n_2 \ra 16 -n_1 - n_2$ the unbroken gauge 
group is mapped to itself, therefore we may restrict 
$2n_2 \leq 16-n_1$ (on the level of the gauge shift this equivalence
is achieved by adding a spinor weight to the shift vector $v$). 
The constraint of modular invariance in four dimensions gives
\equ{
\arry{l}{
2(v^2 - \gf^2) =\frac 18 n_1 + \frac 12 n_2 - \frac 34 \equiv 0. 
}
\labl{LevelMatch4}
}
If this condition is satisfied, the modular invariance requirement
in six dimension $(2 v)^2 - (2 \gf)^2 \equiv 0$ is fulfilled as well,
and hence does not give additional constraints. 
There are two independent set of solutions to this:
\(
(n_1, n_2) = (2+ 8 p_1, 1 + 2 p_2)
\) 
with $p_1 = 0, 1$ and $0 \leq p_2 \leq 3-2p_1$, and 
\(
(n_1, n_2) = (6+ 8 p_1, 2 p_2)
\) 
with $p_1 = 0, 1$ and $0 \leq p_2 \leq 2-2p_1$. This constitutes a
total of ten independent models in four dimensions, while on the six
dimensional hyper surfaces we encounter only a choice of two massless
spectra. 

Let us describe the six dimensional spectrum that correspond to the
orbifold $T^4/\Intr_2$ with gauge shift $2v$. The untwisted matter,
\equ{
\rep{R} ~:~ 2 v \cdot w  - \frac 12 \equiv 0 ~:~ 
( \undr{\pm 1, 0^{n_1-1}}, \undr{\pm 1, 0^{15-n_1}}),
} 
forms the representation $(\rep{2n_1}, \rep{2(16\mbox{-}n_1)})$ of the
six dimensional gauge group. In addition there are two types of
twisted matter 
\equ{
\rep{D}~:~\frac12 (w + v_2)^2 = \mbox{$\frac 14$},  
\quad  \text{or} \quad  
\rep{S}~:~ \frac 12 (w + v_2)^2 = \mbox{$\frac 34$},
} 
where $\tilde v_2 = \frac 12(0^{n_0}, 1^{n_1}, 0^{n_2})$. 
The hyper multiplets $\rep{D}$ have a multiplicity of $20$: 
As these states are obtained from the vacuum by acting with the
oscillator $\ga_{-1/2}^i,\ga_{-1/2}^\ui$ with $i,\ui = 1,2$ which
gives a factor of $2$ when hyper multiplets are counted. As observed
above, $12$ of the $16$ fixed points of $T^4/\Intr_2$ are mapped to
each other, while the $4$ four are inert under the residual $\Intr_4$
action, this gives an additional multiplicity factor of $6+4=10$. 
The states in $\rep{S}$ are pseudo real and form so--called
half--hyper multiplets and they get the same fixed point multiplicity
factor $10$, but count only half. In table \ref{tb:SpecZ4} we give the six
dimensional spectrum.

We turn to the four dimensional spectrum. 
The untwisted matter now comes in two varieties \cite{Aldazabal:1997wi}: 
\equa{
\rep{R}_{i=1,2}~:~ v \cdot w - \frac 14 \equiv 0 ~ : &
\left\{ 
\arry{rl}{
( \rep{2n_0}, \rep{n_1},\rep{1} )_{1} ~: &  
(\underline{\pm 1, 0^{n_1\mbox{-}1}},  \underline{1, 0^{n_1-1}}, 0^{n_2} ),
\\[1ex]
(\rep{1}, \crep{n_1}, \rep{2n_2})_0~: &
(0^{n_0}, \undr{\mbox{-}1, 0^{n_1\mbox{-}1}},  \undr{\pm 1, 0^{n_2-1}}).
}
\right.
\\[2ex] 
\rep{R}_{i=3} ~~~:~ 
v \cdot w - \frac 12 \equiv 0 ~ : & 
\left\{ 
\arry{rl}{
(\rep{1}, \rep{\frac 12 n_1(n_1\mbox{-}1)},\rep{1})_{2} ~: &  
( 0^{n_2}, \underline{1^2, 0^{n_1-2}}, 0^{n_2} ),
\\[1ex]
(\rep{1}, \crep{\frac 12 n_1(n_1\mbox{-}1)}, \rep{1})_{\mbox{-}2} ~: &  
( 0^{n_2}, \underline{\mbox{-}1^2, 0^{n_1-2}}, 0^{n_2} ),
\\[1ex]
(\rep{2n_0}, \rep{1}, \rep{2n_2})_0~: &
(\undr{\pm 1, 0^{n_0-1}}, 0^{n_1}, \undr{\pm 1, 0^{n_2-1}}), 
}
\right. 
}
The former comes with a multiplicity of two, while the latter does not
contribute to anomalies as it consists of vector--like representations
only. In table \ref{tb:SpecZ4} we have collected the untwisted matter
$\rep{R}_{i=1,2}$ that can contribute to anomalies only.

\Zfour{.95}

The next part
of the four dimensional spectrum consists of the zero modes of the six
dimension states. Since only four fixed points of $T^4/\Intr_2$ are
left invariant by the $\Intr_4$ action, only the states on the
corresponding $T^2$ are orbifolded and can give rise to a chiral four
dimensional spectrum. Instead, the six dimensional states on the fixed
points that are mapped to each other, giving a vector--like zero mode
spectrum in four dimensions. In table \ref{tb:SpecZ4} we only give the
chiral spectra that arises from a single fixed point of $T^2/\Intr_2$. The four
dimensional spectrum is completed by four dimensional twisted states,
that are determined by
\equ{
\rep{T}: \frac 12 (w + v)^2 = \frac 3{16}, \frac 7{16}, \frac {11}{16},
}
and are also given in table \ref{tb:SpecZ4}. 
Having determined all possible $\Intr_4$ models and spectra in four
dimensions, we can read off the chiral spectrum from table
\ref{tb:SpecZ4}. (In this table we have not given the vector--like
representations since they are not relevant for anomaly considerations
nor for phenomenology, since they can easily acquire large mass.)

There are two models that may be interesting in the context of GUT
model building: The models with 
$(n_1,n_2)=(2,5)$ and $(10,1)$ contain $\SO{10}$ factors, and both
models contain $\rep{16}$ spinor representations of $\SO{10}$, which
can accommodate full generations of quarks and leptons including
right--handed neutrinos. The models do not have an equal number of
generations because the origin of these spinor representations is
different: The $(2,5)$ models has $16$ generations because the spinors
arise as four dimensional twisted states at the $16$ fixed points of
$T^6/\Intr_4$. The other model is more interesting from the point of
view of phenomenology since it only has four generations. The spinor
$\rep{16}$ is obtained from the orbifolding of the six dimensional 
twisted states that reside at the four fixed points of $T^4/\Intr_2$
that are left inert by the residual $\Intr_4$ action. Both models
suffer from the usual difficulty that the Higgs sector is not rich
enough to give rise to symmetry breaking down to the SM. Further
symmetry breaking can of course be enforced by the inclusion of Wilson
lines and then this model may be a promising starting point for an
orbifold GUT.


\section{Classification of orbifold models}
\labl{sc:ModZ2N}

As the number of $\Intr_3$ and $\Intr_4$ models was still relatively
small the classification could be performed by hand. For arbitrary six
dimensional $\Intr_7$ or $\Intr_{2N}$ ($N=2,3,4,6$) orbifolds 
this becomes a formidable task to be performed by a  
computer, unless some classification systematics is developed both to
identify the modular invariant shifts and to determine the twisted
states.  Here we describe efficient methods to do both and illustrate
them with the $\Intr_3$ and $\Intr_4$ models discussed in section
\ref{sc:ModZ3} and \ref{sc:ModZ4}.  
There are different classes of models depending on the choice of the
spatial shift $\gf$. Moreover, the geometry of fixed hyper surfaces
plays an important role in how the final four dimensional matter
spectrum is composed. Therefore we have organized this section as
follows: First we explain the geometrical structure of the hyper
surfaces within a given $\Intr_{2N}$ orbifold. Next we give a complete
classification of all possible gauge shift vectors. After that we
compute the twisted matter located at the fixed hyper surfaces of the
orbifold. Finally we combine the matter spectra at the various fixed
points to identify the six and four dimensional zero mode
spectrum of the theory.

The geometrical properties of $T^6/\Intr_{2N}$ orbifolds are more
complicated than the prime orbifold $T^6/\Intr_3$, see
\cite{Ibanez:1987pj,Katsuki:1990bf} for a more detailed discussion. 
In a $T^6/\Intr_{2N}$ orbifold we can distinguish the 
hyper surfaces that are left fixed by the orbifold action of 
$\Intr_M \subset \Intr_{2N}$ subgroups by their dimensions. 
The number of entries of the spatial shift $p \gf$ of the $\Intr_M$
subgroup that are non--vanishing modulo one gives the complex
dimensionality. The form of the spatial shift vector $\gf$ of $\Intr_{2N}$
required by supersymmetry implies that these dimensions can be either
four or six. In either case there is a residual $\Intr_{2N}/\Intr_{M}$ action
of the full orbifold group $\Intr_{2N}$ on any hyper surface fixed by
$\Intr_M$. This leads to an identification of all $\Intr_{M}$
hyper surfaces that are mapped to each other by this residual group
action, or to further orbifolding. If the dimension of the hyper
surface is six, a two torus $T^2$ is left inert under the $\Intr_M$
subgroup, and the residual action gives rise to the orbifold
$T^2/(\Intr_{2N}/\Intr_{M})$. The orbifolding of this $T^2$ can be
understood using field theoretical methods as we explain in subsection
\ref{sc:field}.

We restrict our explicit classification resulting in the tables of the
appendices to models with only one subgroup that is $\Intr_2$, 
giving six--dimensional hyper surfaces, to keep
our paper at moderate length. This means that the tables in the
appendices describe the $\Intr_6$--I and $\Intr_8$--I
models, with spatial shifts given in \eqref{StSpaceShift}, only. 
The six dimensional orbifold geometry is always of the form
$T^4/\Intr_2$ with 16 fixed points. The $\Intr_4$ 
orbifold, which we already described in the previous section, indeed
follows these general patterns: The only non--trivial subgroup of
$\Intr_4$ in that case is $\Intr_2$. The twelve of the $\Intr_2$ fixed
tori $T^2$ are mapped each other by the residual $\Intr_2$ action,
leaving 6 independent tori. While the other four fixed points of
$T^4/\Intr_2$ are also fixed points of the $\Intr_4$ 
action giving rise to two--dimensional orbifolds $T^2/\Intr_2$.

Odd order orbifolds can be treated in exactly the same way as the
even order orbifolds when computing the twisted spectra. We show
this by revisiting the $\Intr_3$ twisted states in subsection
\ref{sc:1stZ3}. Only the classification of modular invariant shifts
requires slightly more care. In section \ref{sc:ModZ3} we have
exhausted all possibilities for $\Intr_3$ orbifolds, the $\Intr_7$
case is discussed in appendix \ref{sc:ModZ7}.

\subsection{Classification of modular invariant
$\boldsymbol{\Intr_{2N}}$ shifts}
\labl{sc:shiftClass}

The classification of $\Intr_{2N}$ modular invariant models can be
done on the level of their defining gauge shifts only. 
For a $\Intr_{2N}$ gauge shift we may consider two types of gauge
shift vectors that we refer to as vectorial and spinorial shifts.
For the sake of simplicity we restrict to the vectorial ones only. The
spinorial shifts case can be explored straightforwardly, see
subsection \ref{sc:GenOrbi}. A generic vectorial shift can be brought
to the form    
\equ{
 v = \frac 1{2N}\big( 0^{n_0}, \ldots, N^{n_{N}} \big), 
\quad \text{with} \quad  \sum_{k=0}^N n_k = 16. 
\labl{shift2Nvec}
}
Not all vectorial shift vectors are of this form, but they can be
obtained by adding the vectors $(0^{15},\pm1)$ (or some permutation) 
to $v$ unless not all entries are non--zero. The addition of this
vector results in using the opposite GSO for the twisted states. 
When the shift vector has no zero entries, then changing the sign of
one of the entries to  minus also leads to a different model. 
In this work we ignore these extra possibilities and focus only on the
generic shift vectors as given in \eqref{shift2Nvec}.   
This shift vector lead to the symmetry breaking pattern
\equ{
\SO{32} \ra 
\SO{2n_{0}} \times \U{n_1} \times \ldots \times 
\U{n_{N-1}} \times \SO{2 n_N} 
\labl{Gauge2N}
}
in four dimensions. In case of confusion, like with the $\U{1}$
factors, we employ a subscript to make the distinction between the
various factors, for example 
$\U{n_{\it k}} = \U{1}_k \times \SU{n_{\it k}}$
and $\U{1}_0 = \SO{2 n_0}$ when $n_0 = 1$.

The form of the vectorial  shift vectors introduced above
constitutes a generic choice from a field theoretical point of
view. In string theory only those shift vectors are allowed that lead
to a modular invariant theory, i.e.\ a theory that satisfies the
$\Intr_{2N}$ level matching condition 
\equ{
N (\gf^2 - v^2 ) \equiv 0, 
}
see \eqref{consistency} of appendix \ref{sc:partition}. 
Written in terms of \eqref{shift2Nvec} the level matching condition is
\equ{
N\, \gf^2 \equiv N\, v^2  = 
\frac 1{4 N} \sum_{k=1}^{N} k^2 \, n_k. 
\labl{LevelMatch2Nvec}
}
In the shift vector \eqref{shift2Nvec} we restrict the values of its 
entries shift vector to $0, \ldots, N/(2N)$. This is allowed since the
entries $p/(2N)$ and $(p-2N)/(2N)$ results in the same contribution to
level matching condition: 
\equ{
N \Big(\frac{p}{2N}\Big)^2 \equiv N\Big( \frac{p-2N}{2N} \Big)^2.
}
The signs in the shift vector \eqref{shift2Nvec} are also not relevant
as one can switch the signs of the weight entries correspondingly. 
We can view $v=v_n$ as a function of vector 
$n = (n_1, \ldots,n_{N})$, so that $N \, v^2$ defines a linear
function of $n$ and
\( 
N \, v_{n+n'}^2 = N \, v_n^2 + N \, v_{n'}^2, 
\) 
where the shift $v_{n+n'}$ is defined in the obvious way. We call a
$v_\gn$ a null--shift if $ N \, v_\gn^2 \equiv 0$. Notice 
that if $v_n$ and $v_{n'}$ are solutions of the modular invariance
requirement \eqref{LevelMatch2Nvec}, then $v_{n-n'}$ is a
null--shift. Hence any solution of the level matching condition can
be obtained  as $v_{n+\gn}$ with $v_n$ any fixed solution of
\eqref{LevelMatch2Nvec} and an null--shift $v_\gn$.

This method allows for a simple classification since a particular base
solution of the level matching condition \eqref{LevelMatch2Nvec} is easy
to guess, and the null--solutions are obtained using elementary linear algebra. 
To exemplify  this, we return the $\Intr_4$ case studied in section
\ref{sc:ModZ4}. The null--solutions $v_\gn$ for the level matching
condition \eqref{LevelMatch4} are identified by  
\equ{
\gn = (\gn_1, \gn_2) = (8p_1, 2p_2) + (4, \mbox{-}1)q, 
}
with $p_1, p_2, q$ integers. As base solution we can choose $v_n$ 
with $n=(6,0)$. It is not difficult to see that any solution described
in section \ref{sc:ModZ4} is given by $v_{n+\gn}$. In the appendices
we preform this classification of the modular invariant shift for the
$\Intr_6$--I, $\Intr_8$--I and $\Intr_{7}$ models.

\subsection{Irreducible twisted spectra on
$\boldsymbol{\Cplx^d/\Intr_M}$} 
\labl{sc:spectra}

The next task is to compute the local spectra of matter states
at the fixed points. The untwisted sector can be obtained by
orbifolding the original $\SO{32}$ gauge theory coupled to $\cN=1$
supergravity in ten dimensions. As this can be understood by group
theoretical methods in field theory, it will be postponed to the final
subsection \ref{sc:field} of the present section. Not only the
untwisted sector can be understood using field theoretical
orbifolding, also the orbifolding of the $\Intr_M$ twisted states by
the residual group $\Intr_{2N}/\Intr_{M}$ can be analyzed this
way. For this reason we focus our attention to irreducible twisted
string spectra, i.e.\ spectra that are not obtained by orbifolding
untwisted or twisted states. Moreover, to compute the irreducible
twisted spectra at a given $\Intr_M$ fixed point, it is irrelevant
that the full orbifold is compact or not, this allows us to study the
twisted states on $\Cplx^d/\Intr_M$, which have just a single fixed
point at $0$. By taking the complex dimension
$d = 2, 3$ we can model the fixed points of $T^4/\Intr_M$ and
$T^6/\Intr_M$, respectively. The requirement that we only consider
irreducible string spectra can be translated to the condition that the
$p$th twisted sector of $\Intr_M$ is only taken into account if $p$ is
relatively prime with $M$. Next we move to the technical details of
the classification of twisted states.

For the classification of the irreducible twisted states it is
important to know the chirality of these states. The six dimensional
twisted states form hyper multiplets and therefore have opposite
chirality to that of gauginos, because of the strong constraints of
six dimensional supersymmetry. In four dimensions the chirality of the
fermions in twisted chiral multiplets is both left-- or right--handed. 
For physical investigations, like anomaly considerations, it is
convenient to fix the chirality of all chiral multiplets in the same
way.  The chirality $\gS_p = \pm 1$ of the $p$th twisted sector is
determined by  
\equ{
\frac 14 \gS_p \equiv - \frac 12 \sum_i \tgf_{p\, i},
\labl{chirality}
}
where $\tgf_{p\, i}$ is defined in (\ref{tildeshifts}), see
\cite{Polch2}. The chirality of the $(M\mbox{-}p)$th and the $p$th 
twisted sectors are opposite: $\gS_{M\mbox{-}p} = -\gS_p$. 
Therefore, if we combine the twisted states in chiral multiplets we
only need to take the $p$th or the $(M\mbox{-}p)$th twisted sector
into account. Since the chirality of the untwisted and the first
twisted sector is always positive, we always select the positive
chirality to determine the chiral multiplet representation.

We investigate what kind of representations will be encountered in
heterotic $\SO{32}$ orbifold theories. The irreducible twisted matter
representations is described in terms of tensor products of 
irreducible representations of $\U{n}$ and $\SO{2n}$. Because of 
the two spin--structures of the gauge fermions on the worldsheet
theory we encounter both vectorial and spinorial weights as is
well--known. A derivation of the appropriate mass, GSO and orbifold
conditions is reviewed in appendix \ref{sc:partition}.  The objective
of this section is to give a complete classification of all possible
representations that ever arise, to this end it is extremely useful to
fix form of both the vectorial and spinorial weights to: 
\equ{
\tw = ( 1,\ldots, \mbox{-}1,\ldots, 0,\ldots), 
\labl{StandardWeights}
} 
with all possible permutations. This standard form of the weights
makes general patterns of irreducible representations transparent as
it automatically takes spectral flow into account. To ensure that only
these standard weights \eqref{StandardWeights} can satisfy the
massless condition for the $p$th twisted sector 
\equ{
\frac 12 (\tw + \tv_p)^2 = \frac 58 +  \frac 12 \tgf_p^2 - \tN, 
\qquad 
\tN = \sum_i \Big( \frac 12 - s_i \tgf_{p\, i} \Big) r_i, 
\labl{mass}
}
(where the sum is over all entries of $\gf_i \not \equiv 0$) 
we need that the entries of $\tv_p = \tv^{vec}_p, \tv^{spin}_p$ and
$\tgf_p$ lie between: 
$- \frac 12 < \tv_{p\, I}, \tgf_{p\,i} \leq\frac 12$. 
This requirement uniquely determines the integral vectors 
$d_p^{vec}, d_p^{spin} \in \Intr^{16}$ in the definitions of shifts 
\equ{
\tv^{vec}_p = (pv - d_p^{vec})S_p, 
\qquad 
\tv^{spin}_p = (pv - \frac 12 e - d_p^{spin}) S_p, 
\qquad 
\tgf_p = p \gf + \mbox{$\frac 12$}  e_3 + \gd_p, 
\labl{tildeshifts}
}
where $e = (1^{16})$ and $e_3 = (1,1,1)$. To ensure that all entries
of $\tv^{vec}_p$ are positive and ordered: 
\equ{
\tv_p^{vec} = \frac 1{M} 
\big( 0^{n_0}, \ldots, m^{n_m}\big), 
\labl{StOrdered}
}
where $m = [M/2]$ is the integral part of $M/2$. We have introduced
also the matrix $S_p$ that only has entries equal to $\pm 1$ and $0$. 
(Up to this matrix the vectors $\tv^{vec}_p$ and $\tv^{spin}_p$ are
the same as the shifts $v^{vec}_p$ and $v^{spin}_p$ introduced in
appendix \ref{sc:partition}.) The mass contribution due to
the bosonic oscillators is denoted by $\tN$: The signs $s_i=\pm$ and
integers $r_i  \geq 0$ indicate that $r_i$ bosonic oscillators with internal
space time index $i$ (if $s_i = +$) or $\ui$ (if $s_i = -$) have been
applied to the vacuum state. Not all states that satisfy the mass
condition \eqref{mass} are present in the physical spectrum, only
those states survive that fulfill the GSO and orbifold projections: 
\equ{
\arry{lcrcl}{
\text{GSO} & : & \frac 12 e \cdot \tw 
& \equiv & \frac 12 e \cdot d_p, 
\\[2ex]
\text{Orbifold} & : & 
v S_p \cdot \tw 
& \equiv& \frac 12 (v^2 - \gf^2) p - v S_p \cdot \tv_p
+ {\dsp \sum_i} \gf_i (s_i r_i  + \tgf_{p\,i} ), 
}
\labl{GSOorbi}
}
where in the GSO condition we have used that 
$\frac 12 e \cdot \tw \equiv \frac 12 e \cdot w$. 
The orbifold condition is, in fact, obsolete for the irreducible twisted
sectors, since one can show that by taking the mass condition
\eqref{mass} modulo one and combining it with the GSO projection
implies the orbifold projection.

We need to make one technical comment about the orbifold
phase \eqref{GSOorbi}: If in \eqref{mass} a factor
$\frac12-s_i\tgf_{p\,i} \equiv 0$ for some $s_i$ (let's say $s_i =1$),  
then the corresponding value $r_i$ is meaningless and irrelevant.
This only happens when $\tgf_p \equiv \frac12$, i.e.\ when the twist
$\gf_p$ twist leaves some torus invariant. In this case the twisted
sector is made of six dimensional matter, transforming with a
non--trivial chirality-dependent extra phase under the orbifold
projection (see for example \cite{Nibbelink:2003rc}). In the models
that we consider this argument is only relevant for the $\Intr_2$ twisted
sectors in $\Intr_{2N}$ models, which will be discussed in subsection
\ref{sc:6Dmatter}. We fix the orbifold phases by selecting the same
chirality as the one used in the untwisted sector. In the formalism 
discussed above this can incorporated by setting $r_3$ to some
specific value. In the $\Intr_4$ case $r_3=1$ rather than to $0$. 
Clearly this value of $r_3$ is not due to its original definition of
$r_i$, but it is rather a convenient way of summarizing a 
chirality-dependent phase.

\begin{table}
\begin{center} 
\tabu{|c|c|ccc|c|}{
\hline &&&&& \\[-2ex] 
$\tv$ & group & repr. & weights & prop. & 
mass: $\frac 12 (w + \tv \, e_n)^2$
\\[1ex]\hline\hline &&&&& \\[-2ex]  
$\tv = 0$ & $\SO{2n}$ & $\rep{[2n]}^k$ & 
$(\undr{\pm 1^k, 0^{n-k}})$ & 
$k = 0, 1$ & 
$\frac k2$ 
\\[1ex]\hline &&&&& \\[-3ex]  
$ 0 < |\tv| < \frac 12 $ & $\U{n}$ & $\rep{[n]}_k^\ga$ & 
$\ga (\undr{1^k, 0^{n-k}})$ & 
$\arry{l}{ \ga = \pm \\ k \geq 0}$ & 
$k (\frac 12 + \ga \tv) + \frac n2 \tv^2$ 
\\[1ex]\hline &&&&& \\[-2ex]   
$\tv = \frac 12$ & $\SO{2n}$ & $\rep{2^{n\mbox{-}1}_\ga}$ &
$(\undr{-\frac 12^k, \frac 12^{n-k}}) - \frac 12 e_n$ & 
$\ga = (-)^k$ 
& $\frac {n}8$ 
\\[1ex]\hline
}
\end{center} 
\captn{The twisted matter of $\SO{32}$ orbifold models are
  built out of the following representations: the 
$k$--form representations $\rep{[n]}_k^\pm$ of $\U{n}$ and the vector
$\rep{2n}$ and spinor $\rep{2^{n\mbox{-}1}_\pm}$ 
representations of $\SO{2n}$. 
\labl{tb:mattrep}}
\end{table}

The possible types of gauge groups that arise in heterotic $\SO{32}$
models are either $\SO{2n}$ or $\U{n}$. All $\U{n}$ representations 
are totally anti--symmetric $k$--form tensor representations of 
the vector $\rep{n}$ or its complex conjugate $\crep{n}$, denoted by 
$\rep{[n]}^+_k$ and $\rep{[n]}^-_k$, respectively. (In particular, 
$\rep{[n]}_0^\pm = \rep{1}$, $\rep{[n]}_1^+ = \rep{n}$, 
$\rep{[n]}_1^- = \crep{n}$, and $[\rep{n}]_k=[\crep{n}]_{n-k}$.) 
The representations of $\SO{2n}$ that arise are the fundamental
representation $\rep{[2n]}^k$ or the spinor representation
$\rep{2^{n\mbox{-}1}_\ga}$ of $\ga = \pm$ chirality. 
The index $k=0,1$ is used to simultaneously treat the fundamental and
the singlet representation. In table \ref{tb:mattrep} we have
collected the various representations that can arise and indicate to
which vectorial and spinorial weights they correspond. Moreover, we
have given the value of their mass contribution 
$\frac 12 (w +\tv e_n)^2$ with $e_n=(1^n)$. 
Using these representations we can identify the irreducible 
twisted states for both the vectorial and spinorial
weights. The vectorial weights give rise to representations of the
form  
\equ{
\rep{R}_{vec} = \big( \, 
\rep{[2n_0]}^{k_0}\, ,\,  \rep{[n_1]}^{\ga_1}_{k_1}\, ,\,  \ldots \, , \, 
\rep{[n_{m\mbox{-}1}]}_{k_{m\mbox{-}1}}^{\ga_{m\mbox{-}1}} \, , \, 
\rep{2^{n_m\mbox{-}1}_{\ga_{m}}} \, 
\big),  
\labl{Reprvec}
}
where $\ga_a = \pm$, $k_0=0,1$ and $k_a \geq 0$. The mass
contribution of this state reads 
\equ{
\frac {k_0}2 +  \sum_{a=1}^{m-1} 
k_a \Big(\frac 12 + \ga_a \tv^{vec}_{p\, a} \Big) 
+ \frac 12 (\tv_p^{vec})^2 
=  \frac 58 + \frac 12 \tgf_p^2 - \tN. 
\labl{Massvec}
}
The GSO projection \eqref{GSOorbi} on the vectorial
weights require that 
\equ{
\frac {1-\ga_m}4  +  \frac 12 \sum_{a=0}^{m-1} k_a 
\equiv  \frac 12 e \cdot d_p^{vec}. 
\labl{GSOvec} 
} 
The spinorial weights give very similar representations, except that
the roles of the spinor and vector representations of the ${\rm SO}$
groups are interchanged: 
\equ{
\rep{R}_{spin} = \big(\,  
\rep{2^{n_0\mbox{-}1}_{\ga_0}}\, , \, \rep{[n_1]}^{\ga_1}_{k_1}
\, , \, \ldots, \,
\rep{[n_{m\mbox{-}1}]}_{k_{m\mbox{-}1}}^{\ga_{m\mbox{-}1}}\, , \, 
\rep{[2n_m]}^{k_m}\, 
\big),  
\labl{Reprspin} 
}
where $\ga_a = \pm$, $k_m=0,1$ and $k_a \geq 0$. The mass formula
in this case becomes 
\equ{
\frac{k_m}2 + 
\sum_{a=1}^{m-1} 
k_a \Big(\frac 12 + \ga_k \tv^{spin}_{p\,a} \Big)
+ \frac 12 (\tv_p^{spin})^2 
= \frac 58 + \frac 12 \tgf_p^2 - \tN,  
\labl{Massspin}
}
and the GSO projection \eqref{GSOorbi} on the spinorial
weights reads 
\equ{
\frac {1-\ga_0}4 + \frac 12 \sum_{a=1}^{m} k_a 
\equiv \frac 12 e \cdot d_p^{spin}.   
\labl{GSOspin} 
} 
Let us make some final comments about the irreducible
twisted state representations: Since these are general results, one may
obtain anti--symmetric representation $[\rep{n}]_k$ with $k > n$,
which vanishes identically. We simply drop it all together when it can
never be part of the spectrum. The $\U{1}_a$ charge 
$q_a = (0^{n_0+\ldots}, 1^{n_a}, 0^{\ldots+n_m})$ is computed using 
the formula $q_a S_p\cdot(\tv+\tw)$. For example, the $q_a$ charge of 
the vectorial weight state \eqref{Reprvec} in the first twisted sector
reads 
\equ{
q_a \cdot \rep{R}_{vec} = (\ga_a k_a + n_a v_a )\, \rep{R}_{vec}.
}
This concludes the description of the representations that arise
within a given irreducible twisted sector.

The various twisted sectors are not independent entities, but are
closely related. These relations are encoded in the matrices $S_p$
introduced in \eqref{tildeshifts} above. The defining property of
the matrix $S_p$ is that they bring the gauge shift $p v$ (modulo
integers) of the $p$th twisted sector back to the standard positive
ordered form given in \eqref{StOrdered}. Now assume that the $p$th
twisted sector is irreducible, then the matrix $S_p$ indicates how the
spectrum in this sector can be obtained from the first twisted sector
without any calculation: An off--diagonal entry of this matrix with  
$(S_p)_{ab} = +1$ indicates that in all vectorial and spinorial
representations, \eqref{Reprvec} and \eqref{Reprspin}, of the first
twisted sector one replaces: $\rep{n}_b \ra \rep{n}_a$, while for
$(S_p)_{ab}=-1$: $\rep{n}_b \ra \crep{n}_a$. This specifies the
complete spectrum of the 
$p$th twisted sector up to the spinor chiralities that are determined
by the GSO conditions. By applying the matrix $S_p$ various times one
can generate all irreducible twisted sectors from the first twisted
sector. The number of irreducible twisted sectors is given by the
smallest $K$ such that $(S_p)^K = \Id$. This clearly reduces the
effort of obtaining the irreducible twisted spectra, and shows that
they are intimately related. For example, in appendix \ref{sc:ModZ7} 
we derive the twisted sectors of $\Intr_7$ models. The matrix $S_2$
given in \eqref{SpZ7} satisfies $(S_2)^3 = \Id$, hence all three
twisted sectors can be obtained from the first twisted sector using
$S_2, (S_2)^2$.

The matrix $S_p$ is also defined when the $p$ twisted sector is
reducible, i.e.\ for sectors that are obtained by orbifolding of
sectors of $\Intr_{M/p}$ orbifolds. In this case the matrix $S_p$
indicates how the gauge group in the $\Intr_{M/p}$ model is broken to
the one in the $\Intr_M$ model. Moreover, it indicates whether some
groups and their representations are interchanged or 
complex conjugated. Let us also illustrate this situation
with a concrete example: In section \ref{sc:ModZ4} we saw that the
second twisted sector of $\Intr_4$ models is six dimensional. The
second branching in \eqref{Gauge4} is encoded by the matrix 
\equ{
S_2 = \left( 
\arry{c|c|c}{\Id_{n_0} && \\\hline && \Id_{n_1} \\\hline & \Id_{n_2} &}
\right). 
\labl{S2Z4}
}
In table \ref{tb:SpecZ4} we see the interchange of the
representation, when one goes from the six to the four dimensional
representations. This exemplifies the importance of the matrices
$S_p$, and therefore, in the appendices we will present give them
explicitly for the various orbifold models.

After having derived the general results that determine the  
irreducible twisted spectra, we describe a convenient way to represent
them in the various models. As we will make extensive use of
this representation in the appendices, we illustrate it in subsection
\ref{sc:1stTwZ4}.  For the twisted matter of $\Intr_{2N}$ orbifold
models it proves convenient to employ a combination of three tables to
display the spectrum. The reason for this is that the spectrum is
built out of vectorial and spinorial weights, which receive identical
contributions from the bosonic oscillators through $\tN$ in the mass
formulas  \eqref{Massvec} and \eqref{Massspin}. The first table gives
the possible values of $\tN$, the corresponding index structure of
these states, their ($\SU{2}$) holonomy representations and total
multiplicities.  The other two tables give the possible
representations for vectorial and spinorial weights as functions of
the quantities $\frac 12 \tv^2_{vec}$ and $\frac 12 \tv^2_{spin}$,
respectively. The entries of these two tables are the values of $\tN$
such that the mass condition is fulfilled. The GSO projection leads to
the additional complication, that the chiralities can be model 
dependent. Moreover, for models with $n_0$ or $n_m$ is zero the GSO
conditions \eqref{GSOvec} and \eqref{GSOspin} 
projects out those states that would a have the wrong chirality if the
corresponding spinor would not be zero. All these aspects will be
illustrated by the tables \ref{tb:1stTwT6Z4multi} and
\ref{tb:1stTwT6Z4} of  subsection \ref{sc:1stTwZ4}
where we derive the first twisted spectrum of $\Intr_4$ models.

\subsection{Examples of irreducible twisted spectra}
\labl{sc:ExamplesIrTw}

Because of the generality of the description of the irreducible it
might seem difficult to apply our formalism in concrete situations. 
Therefore we present three examples to illustrate the efficiency of
our method. In the first two examples we determine the first twisted
sectors in the $\Intr_3$ and $\Intr_4$ models that we have studied in
sections \ref{sc:ModZ3} and \ref{sc:ModZ4}, respectively. The
$\Intr_3$ example illustrates how to apply the spectrum formulas
\eqref{Massvec} and \eqref{Massspin}. Instead, we use the $\Intr_4$
case to explain the structure of the tables for twisted states we will
use in our classification of the other $\Intr_{2N}$ and $\Intr_7$
models as well. Our final example determines the universal six
dimensional structure of the $N$th twisted sector in $\Intr_{2N}$
orbifolds.

\subsubsection{First twisted sector in $\boldsymbol{\Intr_3}$ models}
\labl{sc:1stZ3}

\begin{table}
\begin{center} 
\begin{tabular}[t]{rl}
\begin{tabular}[t]{| c | c | c |}
\hline &&\\ [-2ex]
$n$ & vectorial solutions & repr. 
\\[1ex]\hline\hline &&\\ [-2ex]
0 & $r_i = 1, s_i =+, k_a=0$ & $\big(\rep{1}\big)_{\rep{3}}$ 
\\[1ex] &&\\ [-2ex]
& $r_i = 2, s_i =-, k_a=0$ & $\big(\rep{1}\big)_{\crep{6}}$ 
\\[1ex] \hline &&\\ [-2ex]
1 & $r_i=0, k_0=1$ & $\big(\rep{26},\rep{1}\big)_\rep{1}$ 
\\[1ex] &&\\ [-2ex]
& $r_i=0, k_1=3, \ga_1=-$ & $\big(\rep{1}, [\crep{3}]_3\big)_\rep{1}$ 
\\[1ex] &&\\ [-2ex]
& $r_i=1, s_i = -, k_1=1, \ga_1=-$ & $\big(\rep{1}, \crep{3}\big)_{\crep{3}}$ 
\\[1ex]\hline&&\\ [-2ex]
2 & $r_i=0, k_1=2, s_1=-$ & $\big(\rep{1}, [\crep{6}]_2\big)_\rep{1}$ 
\\[1ex] &&\\ [-2ex]
& $r_i=1, s_i=-,k_a=0$ & $\big(\rep{1},\rep{1}\big)_{\crep{3}}$ 
\\[1ex]\hline
\end{tabular} 
& 
\begin{tabular}[t]{| c | c | c |}
\multicolumn{3}{c}{}
\\[0ex] \multicolumn{3}{c}{} \\ [-2ex]
\multicolumn{3}{c}{}
\\[.5ex] \hline &&\\ [-2ex]
3 & $r_i=0, k_1=1, \ga_1=-$ & $\big(\rep{1}, \crep{9}\big)_\rep{1}$ 
\\[1ex]\hline&&\\ [-2ex]
4 & $r_i=k_a =0$ & $\big(\rep{1}, \rep{1}\big)_\rep{1}$ 
\\[1ex]\hline \multicolumn{3}{c}{} \\ [-2ex]
\multicolumn{3}{c}{}
\\[.5ex]\hline&&\\ [-2ex]
$n$ & spinorial solutions & repr. 
\\[1ex]\hline\hline &&\\ [-2ex]
4 & $r_i=0, k_1=0, \ga_0=+$ & $\big(\rep{2^3_+}, \rep{1}\big)_\rep{1}$ 
\\[1ex]\hline&&\\ [-2ex]
5 & $r_i=1, s_i=-,k_1=0,\ga_0=+$ & $\big(\rep{2^0_+},\rep{1}\big)_{\crep{3}}$ 
\\[1ex] &&\\ [-2ex]
& $r_i=0,k_1=1,\ga_1=+, \ga_0=-$ & $\big(\rep{2^0_-},\rep{15}\big)_{\rep{1}}$
\\[1ex]\hline
\end{tabular} 
\end{tabular} 
\end{center}
\captn{\label{tb:SpecT6Z3}
The classification procedure for irreducible twisted states described
in this section is applied to the first twisted sector of $\Intr_3$
models. The subscripts indicate the $\SU{3}$ holonomy representations
of these states. The resulting spectrum for the vectorial and 
spinorial weights coincides with that obtained  in table
\ref{tb:TwStates} by more conventional methods. 
}
\end{table}

The spectrum of the first twisted sector for the $\Intr_3$ orbifold
including the standard representation of the weights were given in
table \ref{tb:TwStates}. Even though we have designed 
the method exposed in the section to be applied to even order
orbifolds, it may also be used for odd order orbifolds, with the
only restriction that $n_m=0$, i.e.\ models with at most a single
${\rm SO}$ group. For the vectorial and spinorial weight states we find that
the mass conditions \eqref{Massvec} and \eqref{Massspin} reduces to  
\equ{
\arry{rrcl}{\dsp 
\text{vec}~:& \dsp 
\frac {k_0}2 + k_1 \Big( \frac 12 + \frac{\ga_1}3 \Big) &=& \dsp 
\frac 16 \Big[ 
4 - n - \sum_i \big( 3 + s_i\big) r_i 
\Big], 
\\[2ex] \dsp \text{spin}~:& \dsp 
k_1 \Big( \frac 12 - \frac{\ga_1}6 \Big) &= &\dsp  
\frac 1{24} \Big[
8 n - 32 - 4 \sum_i \big( 3 + s_i \big) r_i 
\Big]. 
}
} 
We see that the equation for the vectorial weights only has
solutions for $n \leq 4$, while the one for the spinorial weights only
for $n\geq 4$. The solutions have been summarized in table
\ref{tb:SpecT6Z3}, and it is then straightforward to confirm they
correspond to the representations given in table \ref{tb:TwStates} of
section \ref{sc:ModZ3}. All vectorial states that pass the mass shell
condition automatically fulfill the relevant GSO conditions,
\eqref{GSOvec} and \eqref{GSOspin}, as well. For the spinorial weights
the GSO selects the chirality of the spinor representations.

\subsubsection{First twisted sector in $\boldsymbol{\Intr_4}$ models} 
\labl{sc:1stTwZ4}

To illustrate the derivation of the tables we employ in the
appendices, we compute the first twisted sector in $\Intr_4$
orbifolds. The mass conditions for the vectorial and spinorial weights
read 
\equ{
\arry{rrcl}{\dsp 
\text{vec}~:& \dsp 
\frac {k_0}2 + k_1 \Big( \frac 12 + \frac{\ga_1}4 \Big) &=& \dsp 
\frac {11}{16} - \tN - \frac 12 (\tv^{vec}_1)^2, 
\\[2ex] \dsp \text{spin}~:& \dsp 
k_1 \Big( \frac 12 - \frac{\ga_1}4 \Big) + \frac {k_2}2 &= &\dsp  
\frac {11}{16} - \tN - \frac 12 (\tv^{spin}_1)^2, 
}
} 
with 
\equ{
\tN = \frac {r_1 +r_2}4 + \frac {r_3}2,
\qquad 
\frac 12 (\tv_1^{vec})^2 = \frac 1{32} ( n_1 + 4 n_2), 
\qquad  
\frac 12 (\tv_1^{spin})^2 = \frac 1{32} ( n_1 + 4 n_0), 
}
where we have used that $s_1 = s_2 = -$. The possible solutions to 
$\tN = 0, \frac 14$ and $\frac 12$ are given in table
\ref{tb:1stTwT6Z4multi}. The GSO projections determines the
chirality of the spinors:  
\equ{
\text{vec}~:~~ (-)^{k_0+k_1}, 
\qquad 
\text{spin}~:~~ (-)^{n_0+k_0+k_1}, 
}
for the vectorial and spinorial weight representations, respectively. 
The GSO projection for the vectorial weights only depends on the
representation of the twisted states, while for the spinorial weights
it also depends on the model via $n_0$. In table \ref{tb:1stTwT6Z4} we
have given the resulting spectrum for both vectorial and spinorial
weights. We use $\ga=(-)^{n_0}$ to denote the model dependent
chirality of the spinorial weights. By combining the two tables
\ref{tb:1stTwT6Z4multi} and \ref{tb:1stTwT6Z4} the four dimensional
twisted states given in  the last column of table \ref{tb:SpecZ4} are
obtained for the ten four dimensional model tabulated there.

\begin{table}
\begin{center} \begin{tabular}{| c || c | c | c |}
\hline &&&\\ [-2ex]
$\tN$ & $0$ & $\frac 14$ & $\frac 12$ 
\\[1ex]\hline\hline &&&\\ [-2ex]
states & $\ket{0}$ & $\ket{{\undr{1}^k\undr{2}^{1\mbox{-}k}}}$ & 
$\ket{{\undr{1}^k \undr{2}^{2\mbox{-}k}}},  \ket{3}, \ket{{\undr{3}}}$
\\[1ex]\hline&&&\\ [-2ex]
$\SU{2}$ hol. & $\rep{1}$ & $\crep{2}$ & $\crep{3}, \rep{1}, \rep{1}$ 
\\[1ex]\hline&&&\\ [-2ex]
multi. & 1 & 2 & 5 
\\[1ex]\hline
\end{tabular} \end{center}
\captn{
The possible values of $\tN$ for the first twisted sector of a
$\Intr_4$ are $0, \frac 14$ and $\frac 12$. The index structure of the
corresponding states, their $\SU{2}$ holonomy representations
and total multiplicities are indicated. 
\labl{tb:1stTwT6Z4multi}}
\end{table}

\begin{table}
\begin{center} 
\scalebox{.97}{\mbox{
\begin{tabular}{rl}
\begin{tabular}{| l || c | c | c |}
\hline &&&\\ [-2ex]
vectorial repr. \hfill $\backslash$ \hfill $n_1+4 n_2$ &
 \,6\, &14  & 22 
\\[1ex]\hline\hline &&&\\ [-2ex]
$\big(\rep{1},\,\rep{1},\rep{2_+^{n_2\mbox{-}1}}\big)$  &
$\frac 12$ & $\frac 14$ & $0$ 
\\[1ex]\hline&&&\\ [-2ex]
$\big(\rep{1}, \crep{n_1},\rep{2_-^{n_2\mbox{-}1}}\big)$  & 
$\frac 14$ & $0$ &
\\[1ex]\hline&&&\\ [-2ex]
$\big(\rep{2 n_0}, \rep{1},\,\rep{2_-^{n_2\mbox{-}1}}\big)$,  
$\big(\rep{1}, [\crep{n_1}]_{2},\rep{2_+^{n_2\mbox{-}1}}\big)$ &
$0$ & & $\tN$
\\[1ex]\hline
\end{tabular}
& 
\begin{tabular}{| l || c | c | c |}
\hline &&&\\ [-2ex]
spinorial repr. \hfill $\backslash$ \hfill $n_1+4 n_0$ &
\,6\,  & 14   &  22
\\[1ex]\hline\hline &&&\\ [-2ex]
$\big(\rep{\bf 2_{\ga}^{n_0\mbox{-}1}},\rep{1},\rep{1}\big)$  &
$\frac 12$ & $\frac 14$ & $0$ 
\\[1ex]\hline&&&\\ [-2ex]
$\big(\rep{2_{\mbox{-}\ga}^{n_0\mbox{-}1}},\rep{n_1},\rep{\bf 1}\big)$  &
$\frac 14$ & $0$ & 
\\[1ex]\hline&&&\\ [-2ex]
$\big(\rep{\bf 2_{\mbox{-}\ga}^{n_0\mbox{-}1}},\rep{1},\rep{2 n_2}\big)$,   
$\big(\rep{2_{\ga}^{n_0\mbox{-}1}},[\rep{n_1}]_{2},\rep{1}\big)$  &
$0$ && $\tN$ 
\\[1ex]\hline
\end{tabular} 
\end{tabular} 
}}
\end{center}
\captn{
The tables on the left and right display the possible representations 
in the first twisted sector of $\Intr_4$ models for vectorial and
spinorial weights, respectively. The entries of these tables give the
values of $\tN$ which determines the internal space properties of
these states as can be read off from table \ref{tb:1stTwT6Z4multi}. 
The GSO projection selects the chirality of the spinor
representations, which for the spinorial weights depends on 
$\ga = (-)^{n_0}$. When $n_0$ or $n_2$ is zero the GSO only leaves
those states that would have the positive chirality. 
\labl{tb:1stTwT6Z4}}
\end{table}

\subsubsection{Six dimensional matter: the $N$th twisted sector in
$\boldsymbol{\Intr_{2N}}$ orbifolds}
\labl{sc:6Dmatter}

Our final example focuses the $N$th twisted sector which is contained
in all the $\Intr_{2N}$ models. These states live on six dimensional tori
$T^2$ and orbifolds $T^2/\Intr_2$ located at the fixed points of
$T^4/\Intr_2$ embedded inside the orbifold $T^6/\Intr_{2N}$. Starting
from the shift \eqref{shift2Nvec} we find the shift vector of the
$N$th twisted sector  
\equ{
v^{vec}_N = \frac 12 ( 0^{n_0}, 1^{n_1}, 0^{n_2}, \ldots ), 
\qquad 
v^{spin}_N = 
\frac 12 (1^{n_0}, 0^{n_1},1^{n_2}, \ldots ), 
}
using the integral shifts 
\equ{
d_N^{vec} = (0^{n_0+n_1}, 1^{n_2+n_3}, \ldots), 
\qquad 
d_N^{spin} = d_N^{vec} -(1^{n_0}, 0^{n_1}, 1^{n_2}, 0^{n_3}, \ldots).
} 
Even though the signs of $v^{vec}_N$ are all positive, its entries are
not order. With the definitions $m_0 = n_0 + n_2 + \ldots$ and 
$m_1 = n_1 + n_3 + \ldots$, the ordered versions  
\equ{
\tv^{vec}_N = \frac 12 ( 0^{m_0}, 1^{m_1} ), 
\qquad 
\tv^{spin}_N = \frac 12 (1^{m_0}, 0^{m_1}), 
}
are obtained using the matrix $S_N$. (This matrix  $S_2$  for the
$\Intr_4$ case is given in \eqref{S2Z4}.) The mass level and GSO
conditions take very simple forms which are readily solved 
\equ{
\arry{r l l}{
\text{vec}~:~& 
m_1 = 6 - 4 k_0 - 8 \tN, ~ & \dsp
\ga_1 = (-)^{n_2 + n_3 + n_6 + \ldots}, 
\\[2ex] 
\text{spin}~:~& 
m_0 = 6 - 4 k_1 - 8 \tN, ~ & \dsp 
\ga_0 = (-)^{n_0 + n_3 + n_4+ \ldots}. 
}
\labl{SpecT4Z2}
}
Here $\ga_0$ and $\ga_1$ denote the chirality of the spinors for
$k_1=0$ and $k_0=0$, respectively, which are clearly model depend. 
States that contain the vectors of the ${\rm SO}$ groups have spinors
with the opposite chirality as compared to those states without
vectors. The results of this analysis have been collected in table
\ref{tb:SpecTwT4Z2}. The bosonic excitations with $\tilde N=1/2$
generate the states  $\ket{a}$ and $\ket{\undr{a}}$ with $a=1,2$ that
form two doublets under the $\SU{2}$ holonomy group, and the vacuum
$\ket{0}$ with $\tN=0$ is a holonomy singlet. As there are only two
different bosonic representations, we have chosen to collect the
spectrum in a single table. We see that the six dimensional states
(given in table \ref{tb:SpecZ4}) contained in the $T^6/\Intr_4$ models
discussed in section \ref{sc:ModZ4}, in fact, provide the general
structure of matter at the fixed points of $T^4/\Intr_2$.

\begin{table}
\begin{center} 
\begin{tabular}{| l || c | c | c | c|}
\hline &&&&\\ [-2ex]
repr. \hfill $\backslash$ \hfill $m_1$    &
~2~       &
~6~     &
10      &
14
\\[1ex]\hline\hline &&&&\\ [-2ex]
$\big(\rep{1},\rep{2_{\ga_1}^{m_1\mbox{-}1}}\big)$  &
$\ket{a}, \ket{{\undr{a}}}$       &
$\quad \ket{0} \quad $  &
                &
\\[1ex]\hline&&&&\\ [-2ex]
$\big(\rep{2m_0} ,  \rep{2_{\mbox{-}\ga_1}^{m_1\mbox{-}1}}\big)$  &
$\quad \ket{0}\quad $&
&&
\\[1ex]\hline\hline &&&&\\ [-2ex]
$\big(\rep{2_{\mbox{-}\ga_0}^{m_0\mbox{-}1}},\rep{2m_1}\big)$  &
&&&
$\quad \ket{0} \quad $
\\[1ex]\hline&&&&\\ [-2ex]
$\big(\rep{2_{\ga_0}^{m_0\mbox{-}1}},\rep{1}\big)$  &
&&$\quad \ket{0}\quad $&
$\ket{a}, \ket{{\undr{a}}}$
\\[1ex]\hline
\end{tabular} 
\end{center}
\captn{\label{tb:SpecTwT4Z2}
The states on the six dimensional hyper surfaces are determined by 
$m_0 = n_0 + n_2 + \ldots$ and $m_1 = n_2 + n_3 + \ldots$ from the 
$\Intr_{2N}$ shift vector \eqref{shift2Nvec} using the conditions  
\eqref{SpecT4Z2}. The upper two rows correspond to vectorial weight 
representations and the lower two to spinorial weights. The holonomy
doublet states are denoted by $\ket{a}$ and $\ket{\undr{a}}$ 
with $a=1,2$. The chiralities are expressed in terms of 
$\ga_1 = (-)^{n_2 + n_3 + n_6 + \ldots}$ and 
$\ga_0 = (-)^{n_0 + n_3 + n_4+ \ldots}$. 
}
\end{table}

\subsection{The complete string spectrum in six and four dimensions}
\labl{sc:field}

In the previous section we have explained an efficient method to
compute the irreducible twisted states in heterotic $\SO{32}$
orbifold models. With that construction the spectra of these states
were collected in a couple of tables for large classes of models
simultaneously. In this subsection we use such tables to
describe the full massless spectra of $\Intr_{2N}$ models in a field
theoretical language.

For completeness we begin with the gauge sector of the theory. In four
dimensions the $\Intr_{2N}$ orbifold conditions for the ten
dimensional super Yang--Mills theory take the form 
\equ{
A_\gm^w \ra  e^{2 \pi i\, v \cdot w} \, A_\gm^w, 
}
where $w$ are the $\SO{32}$ roots and the shift $v$ is of the form 
\eqref{shift2Nvec}. (The Cartan subalgebra gauge fields $A_\gm^I$
always survive the orbifolding.)  
The four dimensional gauge group therefore is determined by 
$v \cdot w \equiv 0$, its general form is given by \eqref{StOrdered}. 
The untwisted sector matter is obtained from the orbifolding
of the gauge fields with internal spacetime indices
\equ{
A_i^w \ra e^{2\pi i ( w \cdot v - \gf_i)} A_i^w.
}
Their zero mode spectrum is obtained from the condition 
$ w \cdot v - \gf_i \equiv 0$. In addition to these charged states,
there may be neutral untwisted matter that arises from the ten
dimensional supergravity theory.

At the six dimensional hyper surfaces the gauge group and untwisted
spectrum is determined in an analogous way: One only has to replace
$\gf \ra N \gf$ and $v \ra N v$. In fact, the six dimensional gauge
group is always $\SO{2m_0} \times \SO{2m_1}$ where $m_0 =
n_0+n_2+\ldots$ and $m_1 = n_1 + n_3 + \ldots$. 
The common feature of the $\Intr_{2N}$ models that we are focusing on
is that the $N$th twisted sector lives on the six dimensional hyper
surfaces within the orbifold $T^6/\Intr_{2N}$. In the previous
subsection we have used string techniques to determine these six
dimensional states as the twisted sector of $T^4/\Intr_2$. Table
\ref{tb:SpecTwT4Z2} summarizes the resulting spectrum and may be used 
to derive the $N$th twisted spectrum as follows: The value of $m_1$
decides which column is relevant for the spectrum. If the entry of the
table is empty the representation on the left of the corresponding row
is not part of the spectrum. If the entry is $\ket{0}$ the spectrum
contains a holonomy singlet state in the representation determined by
its row. Finally if the entry is $\ket{a}, \ket{\undr{a}}$ the corresponding
representation forms two $\SU{2}$ holonomy doublets. Notice that all
states in table \ref{tb:SpecTwT4Z2} contain spinors. Their chirality
is model dependent and determined by the signs $\ga_0$ and $\ga_1$
defined in the caption. If $m_0$ or $m_1$ is zero there cannot be a
spinor, in that case the state only survives if the would--be
chirality is positive. The spectrum of the six dimensional twisted
matter given in table \ref{tb:SpecZ4} has been determined this
way. Being supersymmetric six dimensional matter, these states are
hyper multiplets. The holonomy singlets form half--hyper multiplets,
i.e.\ hyper multiplets that satisfy a reality condition. This
completes the identification of the six dimensional matter in
$T^6/\Intr_{2N}$ orbifolds.

Also for the four dimensional twisted matter we have produced tables
from which their spectrum can be read off for any specific
$\Intr_{2N}$ model. However, as these spectra can be rather involved, a
combination of three tables have to be employed to identify the
spectrum. Let us explain how one obtains spectra from such tables by
the example of the first twisted sector of a $\Intr_4$ orbifold
discussed in  subsection \ref{sc:1stTwZ4}. In the two tables
\ref{tb:1stTwT6Z4} the possible $\SO{2n_0}\times \U{n_1} \times
\SO{2n_2}$ representations are given. They correspond to two different
types of states in string theory. The chiralities of the spinors in
the table on the left are fixed, while on the right they depend on
value of $n_0$. Like for the $N$th twisted sector described above,
these states are part of the spectrum only if model dependent
quantities ($n_1+4n_2$ and $n_1+4n_0$ in this case) take specific
values given on the top rows of these tables, and only if the
corresponding table entry is not empty. If filled, the value $\tN$ of
the table  entry determines the spacetime properties of the
representation of the corresponding column via table
\ref{tb:1stTwT6Z4multi}: It fixes the internal space index structure
of these states. (From a field theoretical point of view this is
surprising because why should localized states have indices in space
direction to which they cannot propagate.) This is an important 
information since it determines the multiplicities of states, or more
precisely, their $\SU{2}$ holonomy properties. The $\Intr_4$ models
only have one four dimensional sector of chiral multiplets (as the
third twisted sector is the conjugate of the first), but other
$\Intr_{2N}$ model may contain a number of them, as the appendices
show.By providing such a sets of tables for each of them their
spectra are fully specified.

For the irreducible twisted sectors of $\Intr_{2N}$ these tables
to together with transformation matrices $S_p$ that indicate how to
obtain them from the first twisted sector, completely specify their 
four dimensional spectrum. For reducible twisted sectors the situation is
more complicated, as these sectors reside on fixed  points of $\Intr_M$
subgroups of the full orbifold group. These fixed points may
correspond to four or six dimensional spacetime hyper
surfaces. Independently of the dimension, there is a residual action
of $\Intr_{2N}$ on these fixed points, which can have a
multitude of consequences on the spectrum of states on these hyper
surfaces. Whether this leads to identifications, projections, or
further orbifolding of these states depends on the geometrical action
of the residual $\Intr_{2N}$ action only. We now describe the various
possibilities:

As observed at the beginning of this section the residual orbifold
action can lead to the identification of sets of $\Intr_M$ fixed
points. Hence also the twisted matter living on these fixed points
will be identified; no states are projected away. The residual
orbifold action may break the gauge group further. This leads 
to branching of representations of the four dimensional twisted states
with respect to the global gauge group. The chirality of the spectrum
is not lost in this process for four dimensional states. If the
identified fixed points correspond to six dimensional hyper surfaces,
the resulting Kaluza--Klein spectrum in four dimensions is never
chiral.

The other possibility is that the residual $\Intr_{2N}$ action leaves
the $\Intr_M$ fixed point fixed as well. If the dimension of this fixed hyper
surface is four, this leads to a projection of the spectrum at this
fixed point; while if it is six, these states are orbifolded on
$T^2/(\Intr_{2N}/\Intr_M)$. To describe how the spectrum is affected by
this, we introduce some notation: A generic state with
$vect$/$spin$orial weight $\tw$ and $r_i$ internal space indices $i$
(if $s_i=+$) or $\ui$ (if $s_i=-$) in the $p$th twisted sector can be
denoted by  $\ket{^\tw_{s,r}}_p^{vec,spin}$. Note that even the
untwisted matter can be represented like this: $A^w_i = \ket{^w_i}$. 
On such a state the $\Intr_{2N}$ residual action takes the form 
\equ{
\ket{^\tw_{s,r}}_p \ra e^{2\pi i\, \gth_p}
\exp \Big\{{
2 \pi i \big( v S_p \cdot \tw - \sum\limits_i s_i r_i \gf_i \big)
}\Big\} \, 
\ket{^\tw_{s,r}}_p, 
\labl{Orbifolding}
}
as follows from the orbifold condition in \eqref{GSOorbi}. 
The second phase factor can be understood easily in field theory: It
is precisely the transformation property under the orbifold action 
of a state in a representation corresponding to weight $\tw$ that
carries internal space indices parameterized by $s$ and $r$. 
(As explained in section \ref{sc:spectra} and appendix
\ref{sc:partition} if there is an invariant torus then the
corresponding $r_i$ is chosen such that the four dimensional chirality
of these six dimensional orbifold states is the same as that of the
untwisted sector.)
The matrix $S_p$ takes into account that in the classification of
these states, and the resulting tables, $\tw$ refers to weights in the
standard representation given in table \ref{tb:mattrep}. The first
phase factor 
\equ{
\gth_p = \frac 12 (\gf^2 - v^2) p + v S_p \cdot \tv_p - \gf \cdot \tgf_p 
}
has a primarily stringy origin: It results from modular invariance and
spectral flow, see \eqref{tildeshifts}. For six dimensional states
\eqref{Orbifolding} it dictates the orbifold boundary conditions, while
for four dimensional states it defines a projection. In both cases the
four dimensional zero mode spectrum that survives the orbifold
projection satisfies  
\equ{
v S_p \cdot \tw  - {\dsp \sum_i} \gf_i s_i r_i  \equiv 
\frac 12 (v^2 - \gf^2) p - v S_p \cdot \tv_p + \gf \cdot \tgf_p. 
\labl{OrbiProj}
}
This analysis does not take the four dimensional chirality into
account, but this can be obtained straightforwardly 
from \eqref{chirality}. However, if one wants to directly compare to
the untwisted and first twisted spectrum when this chirality is
negative, one needs to take the complex conjugate of the resulting
spectrum. Using these steps the resulting four dimensional spectrum
for any reducible twisted sector can be determined.

Before concluding this section, we would like to return to the
$\Intr_4$ example one final time to illustrate the orbifolding of six
dimensional states. In the fourth column of table \ref{tb:SpecZ4} the
six dimensional spectrum is obtained from table
\ref{tb:SpecTwT4Z2}. The four dimensional spectrum on the fixed points
of $T^2/\Intr_2$ has been determined from the orbifold projection 
\eqref{OrbiProj}. Since the chirality of the second twisted sector
is negative according to \eqref{chirality}, we have conjugated all
representations so that table \ref{tb:SpecZ4} compares only four
dimensional states with the same (positive) chirality.

Let us close this section with a couple of final comments about the 
appendices \ref{sc:ModZ7}, \ref{sc:ModZ6} and \ref{sc:ModZ8} 
the $\Intr_7$, $\Intr_6$ and $\Intr_6$ orbifold models. In these
appendices we 
classify the modular invariant shifts and the irreducible twisted
states only, since they together specify the full heterotic orbifold
model. We refrain from computing the full four dimensional 
spectrum, as that can be obtained using the field theoretical
techniques reviewed in this subsection. As explained with the example
of the first twisted sector of $\Intr_4$ orbifolds in subsection
\ref{sc:1stTwZ4}, we use collection of tables to specify all details
of the irreducible twisted spectra. Moreover, since all irreducible
twisted spectra can be obtained from the first twisted sector, we only
give this spectrum and the matrices that give the other irreducible
twisted states. Presented in this way the spectra of heterotic string
models can be used in a variety of ways. In the next section we mention
a few possible applications of our classification of orbifold models.


\section{Applications and extensions}
\labl{sc:applc}

This section is devoted to some further extensions of our
classifications and possible applications.

\subsection{More general classes of orbifolds}
\labl{sc:GenOrbi}

The explicit classification of models in this work has been restricted
to odd order orbifolds and $\Intr_{2N}$ orbifolds with vectorial
structure, and where only the $N$th twisted sector is six
dimensional. In particular, we have 
neglected the large class of $\Intr_N\times \Intr_{N'}$ models. 
Moreover, we have only focused on vectorial shift vectors. These
omissions have been made for the sake of brevity, rather than as a
matter of principle. In this subsection we take the opportunity to
argue that our methods can be extended without any severe obstacle to
include such models as well.

First of all, for even order orbifolds we may also consider spinorial
shifts\footnote{For odd order orbifold models this simply corresponds
  to interchanging the spin--structures, i.e.\ the interchange of
  vectorial and spinorial weights. The ansatz of \eqref{shift2Nspin}
  can be extended to the most general shift by adding the vectors
  $(0^{15},\pm1)$ or some permutations of them.} 
\equ{
u = \frac 1{4N} 
\big(  1^{n_1},  3^{n_2}, \ldots ,  (2N\mbox{-}1)^{n_N} \big), 
\quad \text{with} \quad 
\sum_{k=1}^N n_k = 16, 
\labl{shift2Nspin} 
}
which give rise to a product of $\U{n}$ gauge groups only 
\equ{
\SO{32} \ra 
\U{n_1} \times \U{n_2} \times \ldots \times \U{n_{N}}. 
\labl{GaugeSpin}
}
The classification of models with vectorial and spinorial shift
vectors looks different. The spinorial shifts
\eqref{shift2Nspin} can be classified in much the same way as the
vectorial ones in subsection \ref{sc:shiftClass}. In this case the
modular invariance condition gives the linear equation  
\equ{
N\, \gf^2 \equiv N\, u^2  = 
\frac 1{16 N} \sum_{k=1}^{N} (2k-1)^2 \, n_k. 
\labl{LevelMatch2Nspin}
} 
As this can be treated as a linear system for the numbers $n_k$, 
the same method of null--solutions may be applied as described in
subsection \ref{sc:shiftClass}.

Not only can the spinorial shift vectors be classified, also our
method of systematically determining all twisted states applies with
only a few minor modifications. The allowed representations in the
irreducible twisted sectors are still determined by the mass
conditions, \eqref{Massvec} and \eqref{Massspin}, except that, since
the resulting gauge group \eqref{GaugeSpin} never contain $\SO{2n}$
groups, there are no $k_0$ or $k_m$ contributions. For the same reason
the GSO conditions, \eqref{GSOvec} and \eqref{GSOspin}, are now always
true projections, rather than selection rules for the chirality
of spinor representations.

Also the extension of our classification method to $\Intr_N \times \Intr_{N'}$
orbifolds is straightforward. This class of orbifolds also includes 
$\Intr_{2N}$ orbifolds for which the $N$th twisted sector is not a six
dimensional sector on the orbifold $T^4/\Intr_2$. A
$\Intr_N\times\Intr_{N'}$ orbifold is defined by two spacetime shifts,
$\gf$ and $\gf'$, and two gauge shifts, $v$ and $v'$. The requirements
of modular invariance are \cite{Ibanez:1987pj,Forste:2004ie}
\equ{
\frac 12 {N_{p\,p'}} \Big( 
\big(p\gf+p'\gf' \big)^2 - \big(pv+p'v'\big)^2 
\Big) \equiv 0, 
\labl{ModInvZNZN'}
}
for all $p,p'$, where $N_{p\,p'}$ is the order of the shift $p\gf+ p'\gf'$.
Using similar techniques as employed in subsection
\ref{sc:shiftClass}, all solutions can be determined by some linear
algebra. For a $\Intr_N$ orbifold the signs of the shift vector can
always be rotated away, but this need not be the case anymore because
of the third condition in \eqref{ModInvZNZN'}.  The $(N N'\! -  1)$
twisted sectors are labeled by two integers $p= 0,\ldots, N-1$
and  $p'=0,\ldots, N'\! -1$ not both equal to zero. To decide whether
the $(p,p')$ sector is four or six dimensional, one computes the
relevant spacetime shift $p\gf+p'\gf'$. For each sector we may define
the generalization of \eqref{tildeshifts} by 
 \equ{
\arry{l}{
\tv^{vec}_{p\,p'} = (pv+p'v' - d_{p\,p'}^{vec})S_{p\,p'}, 
\\[2ex]
\tv^{spin}_{p\,p'} = (pv+p'v' -  \mbox{$\frac 12$}e -
d_{p\,p'}^{spin}) S_{p\,p'}, 
}
\qquad 
\tgf_{p\,p'} = p \gf+p'\gf' + \mbox{$\frac 12$}  e_3 + \gd_{p\,p'}.
\labl{tildeshiftsZNZN'}
}
Hence the spectrum classification formulas \eqref{Massvec} and
\eqref{Massspin} and GSO projections \eqref{GSOvec} and
\eqref{GSOspin} may be used to determine all irreducible twisted
states in each $(p,p')$ sector.

Finally, also the extension to models with Wilson lines can be
performed without difficulty. As it is well--known also modular
invariance put stringent conditions on the possible Wilson lines 
\cite{Ibanez:1987pj}, which can be analyzed in a similar fashion as
the multiple shift vectors for $\Intr_N\times \Intr_{N'}$
orbifolds. For the twisted spectrum the situation is fully identical
to the one studied at length in this work, since, as we used in
subsection \ref{sc:spectra}, the twisted states are determined by the
local shift vector only. In theories with Wilson lines not all fixed
points are equivalent, but to each fixed point a local shift vector is
associated which determines the complete (irreducible twisted)
spectrum at this fixed point \cite{Gmeiner:2002es}. Hence, for models
with Wilson lines our method of computing spectra is very efficient.

\subsection{Applications to the $\boldsymbol{\E{8}\times\E{8}}$
theory}
\labl{sc:ApplcE8}

The heterotic $\E{8}\times\E{8}'$ theory has been studied much in the past since
this string theory was the first that looked promising for
phenomenology. However, to go beyond identifying the gauge group and
untwisted states always proved difficult because it was impossible to
recognize patterns in the twisted states. One of the reasons for this
is the appearance of various exceptional groups, whose 
representations do not follow easily identifiable patterns.   
The method that we have used to classify the twisted states of the
heterotic $\SO{32}$ string on orbifolds in subsection \ref{sc:spectra}
can be extended to the $\E{8}\times\E{8}'$ theory, as we will now
demonstrate. The central observation is to classify the subgroups and
their representations of the maximal subgroup $\SO{16}\times\SO{16}'$
rather than those of the $\E{8}\times\E{8}'$ group itself.

Using elementary representation theory which can be found in the
tables of \cite{Slansky:1981yr} the identification of
representations of exceptional groups is not difficult. For example,
as is well--known, the group $\E{8}$ can be understood as the spinor
bundle over $\SO{16}$. This exemplifies that the representations of
$\E{8}$ and other subgroups can be easily understood as combinations
of representations of (maximal) regular subgroups.

In an $\SO{16}\times\SO{16}'$ Cartan basis a gauge shift is now
described by the combination $(v, v')$ of two shift vectors. In general
each of the $\SO{16}$ groups is broken to 
\equ{
\SO{16} \ra \SO{2n_0} \times \U{n_1} \times \ldots \times 
\U{n_{N\sm1}}\times \SO{2n_N}.
}
We may denote the $\SO{16}\times\SO{16}'$ weights as $(w,w')$, which 
can independently be vectorial or spinorial weights. By using the
definitions \eqref{tildeshifts} we bring the weights in the standard
form of table \ref{tb:mattrep}, and therefore the states can be
classified as before. In particular, the mass formulas are now
split into four sectors:
\equ{
\arry{c rcl}{
(\rep{vec},\rep{vec}') ~:& 
N_{vec} + N_{vec}' & = & \frac 58 + \frac 12 \tgf_p^2 - \tN, 
\\[2ex] 
(\rep{vec},\rep{spin}') ~:& 
N_{vec} + N_{spin}' & = & \frac 58 + \frac 12 \tgf_p^2 - \tN, 
\\[2ex] 
(\rep{spin},\rep{vec}') ~:& 
N_{spin} + N_{vec}' & = & \frac 58 + \frac 12 \tgf_p^2 - \tN, 
\\[2ex] 
(\rep{spin},\rep{spin}') ~:& 
N_{spin} + N_{spin}' & = & \frac 58 + \frac 12 \tgf_p^2 - \tN, 
}
\labl{E82mass}
}
where we have used the shorthand notations 
\equ{
\arry{rcl}{
N_{vec} &=& \frac {k_0}2 +  \sum\limits_{a=1}^{m-1} 
k_a \Big(\frac 12 + \ga_a \tv^{vec}_{p\, a} \Big) 
+ \frac 12 (\tv_p^{vec})^2,  
\\[2ex] 
N_{spin} &=& 
\frac{k_m}2 + 
\sum\limits_{a=1}^{m-1} 
k_a \Big(\frac 12 + \ga_k \tv^{spin}_{p\,a} \Big)
+ \frac 12 (\tv_p^{spin})^2, 
} 
}
for both $\SO{16}$ and $\SO{16}'$ weights. The solutions to the mass
relations \eqref{E82mass} can be solved as quickly as the ones given
in \eqref{Massvec} and \eqref{Massspin} of the $\SO{32}$ theory. 
On both $\SO{16}$ weights the GSO projections, \eqref{GSOvec} and
\eqref{GSOspin}, are applied.

As our discussion here showed, our classification method can be applied
directly to the $\E{8}\times\E{8}'$ theory as well, and will also in
that case give fast classification results. More importantly, it will
also make the patterns in the twisted spectra transparent.

\subsection{Heterotic/Type--I duality on odd order orbifolds}
\labl{sc:hettypI}

The classification procedure that we describe in this paper might be
useful in the context of the $S$--duality \cite{Polchinski:1995df}
between the heterotic  
$\SO{32}$ string and type--I models. We illustrate this by make some
comments on the present status of this duality in four dimensions. 
We begin by briefly recalling the construction of type--I models. 
The starting point is type--II closed string theory with an operator
$\gO$ that reverse the spatial worldsheet coordinate. Keeping only the
invariant closed strings implies that at tree level the worldsheet
cylinder has become a strip $\Real \times S^1/\Intr_2$, and 
hence it is natural that this theory includes open strings as well. 
This introduces in the model non--dynamical orientifold planes that are 
sources for $RR$--flux. If the internal space is compact,
consistency of this construction is enforced by requiring that  
all these tadpoles are canceled which requires specific sets of
$D$--branes \cite{Polchinski:1995mt}. (In a rather different 
but earlier approach the construction was also described in 
\cite{Pradisi:1988xd}). 
A stack of $D$--branes gives rise to $\U{n}$, $\Sp{2n}$ or $\SO{2n}$
gauge groups generated by their Chan--Paton labels \cite{Witten:1997bs} 
(see also \cite{deBoer:2001px}). 
The tadpole cancellation conditions is powerful enough to ensure that 
irreducible anomalies do not arise in any dimension
\cite{Aldazabal:1999nu}. In particular in ten dimensions the type--I
theory is required to be an $\SO{32}$ gauge theory due to the
32 $D9$--branes.

In ten dimensions the resulting type--I supergravity is described by the
same action as the heterotic $\SO{32}$ supergravity but with a 
different coupling. On this observation the strong/weak duality
between these two theories on the supergravity level was based, and
then extended as a string duality in ten dimensions in
\cite{Polchinski:1995df}.  Ref.\
\cite{Angelantonj:1996uy,Aldazabal:1998mr} consider the duality
relation between the dilatons of the two theories in various
dimensions by toroidal compactification. In particular, these authors
observe that in four dimension the duality is a weak/weak duality in
the supergravity approximation. For more complicated compactification,
like orbifolds, the status of the duality is less clear. In
particular, the consistency requirement of modular invariance of the
heterotic string, which enforces the existence of twisted states, does
not have a counterpart for the open string in type--I. 
In this subsection we investigate the four dimensional version of the
heterotic/type--I $S$--duality by matching gauge group and chiral
spectra of four dimensional heterotic $\SO{32}$ orbifold models to
those of type--I orbifolds.

The details of the classification on the type--I side was established by
\cite{Angelantonj:1996uy,Aldazabal:1998mr}, where also previous results 
were collected, we refer to these papers for a detailed list of references.
Concretely, the tadpole cancellation conditions impose that 
\begin{equation}
\Tr\big[ \gamma^{2k} \big] = 32 \prod_{i=1}^3 \cos(\pi k \phi_i), 
\quad \text{with} \quad 
\gamma=e^{-2 i \pi v\cdot H}, 
\end{equation}
for $0 \leq k \leq N- 1,$ where the Cartan generators $H_I$ act on
the $SO(32)$ Chan--Paton factors. These conditions fix the shift $v$
uniquely. In particular, the $\Intr_3$ orbifold of type--I theory has 
$v = (0^4, 1^{12})/3$, which gives the same gauge group and charged
untwisted matter fields as in the heterotic $\Intr_3$ model with
$n=4$, see section \ref{sc:ModZ3}. Only the
untwisted states give rise to a charged spectrum in type--I models,
since these models cannot contain charged twisted states. This gives a
partial explanation why there is one $\Intr_3$ type--I model: As can
be seen from \eqref{unAnom} only for $n=4$ the dual heterotic theory
does not have irreducible anomalies in its untwisted spectrum. 
But anomaly cancellation does not provide a complete answer, since on
the  heterotic side also the $n=0$ model is obviously anomaly free, but
it has no type--I counterpart. The situation is identical for the
other odd orbifold $\Intr_7$: From the untwisted spectrum of the
heterotic $\Intr_7$ models given in appendix \ref{sc:ModZ7} it follows
that two model are free of irreducible anomalies, see
\eqref{IrrAnomZ7}. One has the shift vector $v = (0^4,1^4,2^4,3^4)/7$;
for this shift a type--I  model exists, see \cite{Aldazabal:1998mr},
while the trivial embedding ($v=0$) again has no type--I dual.

There are more aspects that need to be addressed to establish how
the $S$--duality is a true duality between type--I and heterotic
theories in four dimensions. First of all, also the other heterotic
$\Intr_3$ models should have duals  on the type--I side. Such dual
models require additional four dimensional states to appear in order
to cancel the irreducible anomaly. Maybe these 
'twisted open string states' can be understood as $D$--string
excitations \cite{Polchinski:1996fm}.

Moreover, even for type--I and
heterotic models with equivalent $\Intr_3$ and $\Intr_7$ shifts, the
spectra of these dual theories are not equal. It has been argued in 
\cite{Kakushadze:1997wx,Kakushadze:1998cd} the $\Intr_3$ twisted
states become massive when the orbifold singularities are blown up,
and then the remaining massless charged spectra match. Another 
profound difference between these two models is that the cancellation
of the leftover reducible anomalies is fundamentally different for heterotic
and type--I models. In heterotic models the Green--Schwarz mechanism
always descents from the ten dimensional one, and therefore requires a
unique factorization. As we saw in section \ref{sc:ModZ3} for this
factorization the charged twisted states  are essential, see
\eqref{fac4Dhet} for example. Instead, in type--I orbifold models, it
is sufficient for local anomaly cancellation that the anomaly
polynomial factorizes, as the anomalous couplings of the neutral
twisted closed particles at the fixed points are sufficiently flexible
\cite{Ibanez:1998qp,Morales:1998ux,Scrucca:1999zh}.

In the discussion above we have primarily focused on the two odd
order orbifolds, therefore one may wonder what the status is of the
heterotic/type--I duality on even order orbifolds. As was first
observed in ref.\ \cite{Gimon:1996rq}, even order orbifolds 
require that  both $D9$-- and $D5$--branes are introduced. This system
has both gauge groups in ten and six dimensions. This construction has
been extended to four dimensions by various groups 
\cite{Dabholkar:1996ka,Blum:1996hs,Klein:2000hf,Klein:2000qw}.
In particular, the authors of ref.\ \cite{Aldazabal:1998mr} showed
that there are no $\Intr_4$ type--I models, that fulfill the tadpole
cancellation conditions. As we have seen in section \ref{sc:ModZ4},
there exist ten perturbative heterotic $\Intr_4$ models, but there are
apparently no possible type--I duals. The gauge groups localized on
the six dimensional $D5$--branes do not have perturbative counter
parts on the heterotic side, but might be related to non--perturbative
$M5$--brane excitations. The $M5$--branes give rise to non--modular
invariant heterotic models. Part of their spectrum can sometimes be
matched to that of type--I models \cite{Aldazabal:1997wi}. The duality
between heterotic and type--I four dimensional models in general, and
for even order orbifolds in particular, still requires further research. 
We hope that our classification of heterotic $\SO{32}$ 
models may provide a useful testing ground for new proposals for more
precise definitions of this duality.

\subsection{Model searches}
\labl{sc:searches}

As a final application we mention that our classification procedure
can be very useful for string model searches. This can be searches for
MSSM--like, GUT or orbifold GUT models. Irrespectively of which kind
of model one is looking for, the basic strategy is the same: First
find a model with the appropriate gauge group, secondly check whether
at least the wanted matter representations are present, next fill in
the details of the spectrum. After this, more detailed investigations
can be undertaken in which the forms of (perturbative)
superpotentials, gauge kinetic terms and \Kh\ terms are obtained.

Our methods can clearly be aimed at tackling the first part of this
program, as it only relies on the spectrum of the theory. Given the
gauge group one is looking for, one can immediately select possible
orbifolds models that have that gauge group unbroken. Since we have
made the connection between the resulting gauge groups and the gauge
shifts explicit in \eqref{shift2Nvec} and \eqref{Gauge2N}, we can
identify the appropriate values of some of the integers $n_i$ in the
gauge shift. The requirement of modular invariance then quickly tells
us if solutions can be found at all. Using our tables one has 
immediate overview of the spectrum of the theory.

Let us close this section on applications by illustrating how model
searches can be performed using our classification. Suppose one is
interested in obtaining $\SO{10}$--like GUTs from string theory. For
the even order $\Intr_{2N}$ orbifolds, $\SO{10}$ groups arise if
$n_0=5$ or $n_N=5$. Moreover, since we have shown that the six and
four dimensional twisted sectors contain spinor representations, it is
quite likely that the chiral string spectrum contains spinors of
$\SO{10}$. In particular if we focus on the $\Intr_4$ model, it
follows from the modular invariance condition that $n_1=2,10$ for
either $n_0=5$ or $n_2=5$. These solutions correspond precisely to the
two models found in section \ref{sc:ModZ4} from table
\ref{tb:SpecZ4}. (the other two are the same as these ones as they are
obtained by interchanging the roles of $n_0$ and $n_2$.)


\bibliographystyle{paper}
{\small
\renewcommand{\bibsep}{-.5ex} 
\bibliography{paper}
}

\newpage 
\appendix 
\def\theequation{\thesection.\arabic{equation}} 
\setcounter{equation}{0}

\section{Non--compact holomorphic partition function}
\labl{sc:partition}

In this appendix we collect some properties of the heterotic $\SO{32}$
string partition function on which the analysis of the main text
relies heavily. The partition function is constructed out of the
following modular forms:
\equ{
\gvth\brkt{\ga}{\gb} (\gt) 
= 
e^{ 2 \pi i \ga \gb}\,  q^{\frac 12 \ga^2} \, 
\prod_{n \geq 1} \Bigl\{
\bigl( 1 - q^n \bigr) 
\prod_{s= \pm} 
\bigl( 1 + e^{-2 \pi i s \gb}\, q^{n - \frac 12 - s \ga} \bigr) 
\Bigr\},
\quad 
\get(\gt) = q^{\frac 1{24}} 
\prod_{n \geq 1} \bigl( 1 - q^n \bigr), 
\labl{thetaeta}
}
with $q = e^{2\gp i\, \gt}$.
In the sum representation the theta function can be easily generalized
to vector valued characteristics. In particular for the heterotic
$\SO{32}$ theory with spin--structures $t,t'$ and $p, p'$ at one loop
we use the theta function 
\equ{
\gvth\brkt{\frac {1-t}2 \,e - p\, v }{\frac{1-t'}2e -p'v}(\gt) = 
\sum_{n \in \Intr^{16}} q^{\frac 12 (n + \frac {t-1}2 e + pv )^2}
e^{2\pi i (\frac {t'-1}2e +p' v )\cdot (n + \frac {t-1}2 e + p v)},
\labl{lambdatheta}
}
such that the full holomorphic partition function can be written as 
\equ{
Z = \frac 1{2N} \sum \tget^{p\,,t\,}_{p',t'} \, 
\frac 1{\get^{15}(\gt)}
\gvth\brkt{\frac {1-t\,}2\, e - p \, v }{\frac{1-t'}2 e -p'v}(\gt)
\Big( 
\prod_i \gvth\brkt{\frac 12 + p\, \gf_i}{\frac 12 + p' \gf_i}(\gt) 
\Big)^{-1}, 
\labl{partition}
} 
where $N$ is the order of the orbifold. The phases 
\equ{
\tget^{p\,, t\,}_{p',t'} = \exp 2 \pi i \Big(  
\frac 12 ( \gf^2 - v^2 ) 
\, p p' 
+ p \frac {1-t'}2 e \cdot v
\Big) 
}
are determined by the requirement of modular invariance. This
requirement also implies that 
\equ{
\frac N2 ( v^2 - \gf^2) \equiv 0. 
\labl{consistency} 
}

The zero mode spectrum of the theory is determined by expanding the
partition function \eqref{partition} to the constant part in $q$. 
This expansion can be facilitated by first determining the minimal
power that can arise from the theta function is the numerator and
denumerator by using the periodicity of the theta functions in their
upper characteristic. The product representation 
\equ{
\gvth\brkt{\frac 12 + p\, \gf_i}{\frac 12 + p' \gf_i} = 
e^{2 \pi i \tgf_{p\, i}(p' \gf_i + \frac 12)} 
q^{\frac 12 \tgf^2_{p\,i}} 
\prod_{m_i \geq 1} \Big\{ 
( 1 - q^{m_i} ) \prod_{s_i = \pm} 
\Big( 1 + e^{-2\pi i s_i(p' \gf_i + \frac 12)} 
q^{m_i - \frac 12 - s_i \tgf_{p\, i}} \Big) 
\Big\}
\labl{bospart}
} 
is convenient when expanding 
$\gvth\brkt{\frac 12 + p\, \gf_i}{\frac 12 + p' \gf_i}^{-1}$: 
since all terms with $m_i > 1$ give massive string states and can be
ignored since we are only interested in the massless spectrum. 
Here we have defined the vector $\gd_p\in \Intr^3$ such that all entries 
\( 
\tgf_p = p \gf + \frac 12 e + \gd_p, 
\) 
lie between $- \frac 12 < \tgf_{p\, i} \leq \frac 12$. 
We rewrite \eqref{lambdatheta} for $t = 0$ as 
\equ{
\gvth\brkt{\frac 12\, e -p\,v}{\frac {1-t'}2e - p'v} 
= \sum_{w \in \Intr^{16}} 
q^{\frac 12 ( w + v^{spin}_p)^2} 
e^{2\pi i ( \frac {t'-1}2 + p' v)(w+ v^{spin}_p)}, 
}
where $d_p^{spin} \in \Intr^{16}$ is chosen such that all entries 
\(
v^{spin}_p = p v - \frac 12 e - d_p^{spin}, 
\) 
lie between $- \frac 12 < v^{spin}_{p\, I} \leq \frac 12$. 
Similarly for the vectorial weights ($t=1$) we have 
\equ{
\gvth\brkt{-p\,v}{\frac {1-t'}2e - p'v} 
= \sum_{w \in \Intr^{16}} 
q^{\frac 12 ( w + v_p^{vec})^2} 
e^{2\pi i ( \frac {t'-1}2 e + p' v)(w+ v_p^{vec})},
}
with $d_p^{vec} \in \Intr^{16}$ also chosen such that the entries of 
\(
v^{vec}_p = p v - d_p^{vec}
\) 
lie in the same interval. 
The zero mode mass spectra are determined by the relations 
\equ{
\frac 12 (w + v_p)^2 - \frac 58 - \frac 12 \tgf_p^2 + 
\sum_i \Big( \frac 12 - s_i \tgf_{p\, i}\Big) r_i = 0, 
\labl{massA}
}
with $s_i=\pm$ and integers $r_i  \geq 0$ and 
$v_p = v_p^{vec}, v_p^{spin}$ for vectorial and spinorial
weights, respectively. 
Moreover it can be recognized that the sums over $t'$ and $p'$
lead to GSO and orbifold projections that lead to the conditions
\equ{
\arry{lcl}{
\text{GSO} & : & \frac 12 e\cdot ( w - d_p) \equiv 0, 
\\[2ex] 
\text{Orbifold} & : & 
\frac 12 (\gf^2 - v^2) p + v\cdot (w + v_p) - 
{\dsp \sum_i} \gf_i ( \tgf_{p\, i}  + s_i r_i ) \equiv 0, 
}
\labl{phaseGSOA}
}
on the spectrum for both the vectorial and spinorial weights. 
The integer number $r_i$ in \eqref{massA} is only relevant 
when $ \frac12-s_i \tgf_{p\,i} \not \equiv 0$. If a $p$th twisted sector
completely fills a torus $T^2$, then in the reduction to four
dimensional the phase \eqref{phaseGSOA} is affected by an extra
chirality--dependent term. The value of this phase is only relevant in
this paper for the $\Intr_4$ models presented in table \ref{tb:SpecZ4}
given section \ref{sc:ModZ4}. The appropriate phase can be obtained by
setting taking $r_3=1$ in \eqref{phaseGSOA}.




\newcommand{\Zseven}{
\def\thetable{\ref{sc:ModZ7}.\arabic{table}} 
\setcounter{table}{0}
\begin{table} 
%
%
\begin{center} 
\[
\arry{| l | l |}{
\hline & \\[-2ex]  
\Intr_{7} & \gf = \frac 1{7}\big(1,2, \mbox{-}3 \big)
\\[1ex] \hline \hline & \\[-2ex] 
\gf_{1} = \frac 1{7} & 
[\crep{n_3}]_2 , 
\big(\rep{2n_0}, \rep{n_1}   \big) , 
\big( \crep{n_1}, \rep{n_2}  \big) , 
\big(  \crep{n_2}, \rep{n_3} \big) 
\\[1ex] \hline & \\[-2ex] 
\gf_{2} = \frac 2{7} & 
[\rep{n_1}]_2 , 
\big(\rep{2n_0}, \rep{n_2}   \big) , 
\big( \crep{n_1}, \rep{n_3}  \big) , 
\big(  \crep{n_2}, \crep{n_3} \big) 
\\[1ex] \hline & \\[-2ex] 
\gf_{3} = \mbox{-}\frac 3{7} & 
[\rep{n_2}]_2 , 
\big(\rep{2n_0}, \crep{n_3}   \big) , 
\big( \crep{n_1}, \crep{n_2}  \big) , 
\big(  \rep{n_1}, \rep{n_3} \big) 
\\[1ex] \hline 
}
\]
\end{center}
\captn{
This table gives untwisted spectrum of the $\Intr_{7}$ orbifold models.  
\labl{tb:SpecUnZ7}}
%
%
\begin{center} 
\begin{tabular}{| c || c | c | c | c | c |}
\hline &&&&&\\ [-2ex]
$\tN$ & $0$ & $\frac 17$ & $\frac 27$ & $\frac 37$ & $\frac 47$
\\[1ex]\hline\hline &&&&&\\ [-2ex]
states          & 
$\ket{0}$       & 
$\ket{\undr{1}}$ & 
$\ket{\undr{1}^2}, \ket{\undr{2}}$ & 
$\ket{\undr{1}^3}, \ket{\undr{1}\undr{2}}, \ket{3}$ & 
$\ket{\undr{1}^4}, \ket{\undr{1}^2\undr{2}}, \ket{\undr{2}^2},
\ket{\undr{1}3},\ket{\undr{3}}$ 
\\[1ex]\hline&&&&&\\ [-2ex]
multi. & 1 & 1 & 2 & 3 & 5 
\\[1ex]\hline
\end{tabular} 
\end{center} 
\captn{
The possible values of $\tN$ for the first twisted sector of a
$\Intr_7$ are $0, \frac 17, \ldots \frac 47$. The index structure of the
corresponding states and total multiplicities are indicated. 
\labl{tb:1stTwZ7Imulti}}
%
%
\begin{center} 
\begin{tabular}{| l | l  || c | c | c | c | c |}
\hline \multicolumn{2}{|c||}{}&
\multicolumn{5}{|c|}{}
\\ [-2ex]
\multicolumn{2}{|l||}{vectorial repr.} &
\multicolumn{5}{|c|}{}
\\[1ex] \hline &&
\multicolumn{5}{|c|}{}
\\ [-2ex]
even \hfill $\backslash$ \hfill $n_1+4 n_2+9n_3$ &
odd \hfill $\backslash$ \hfill $n_1+4 n_2+9n_3+7$& 
14            &
28           &
42           &
56           &
70
\\[1ex]\hline\hline &&&&&&\\ [-2ex]
$\big(\rep{1}\big)$ &
$\big(\crep{n_3}\big)$&
${}\frac{4}{7}$&
${}\frac{3}{7}$&
${}\frac{2}{7}$&
${}\frac{1}{7}$&
${}0$  
\\[1ex]\hline&&&&&&\\ [-2ex]
$\big([\crep{n_3}]_2\big)$&
$\big(\crep{n_2}\big)$,$\big([\crep{n_3}]_3\big)$&
${}\frac{3}{7}$&
${}\frac{2}{7}$&
${}\frac{1}{7}$&
${}0$         &

\\[1ex]\hline&&&&&&\\ [-2ex]
$\big(\crep{n_2},\crep{n_3}\big)$,$\big([\crep{n_3}]_4\big)$&
$\big(\crep{n_1}\big)$,$\big(\crep{n_2},[\crep{n_3}]_2\big)$&
${}\frac{2}{7}$& 
${}\frac{1}{7}$& 
${}0$         &
                      &
\\[1ex]\hline&&&&&&\\ [-2ex]
$\big(\crep{n_1},\crep{n_3}\big)$,
$\big(\crep{n_2},[\crep{n_3}]_3\big)$ \qquad\qquad\qquad   &
$\big(\rep{2n_0}\big)$,
$\big(\crep{n_1},[\crep{n_3}]_2\big)$,
$\big([\crep{n_2}]_2,\crep{n_3}\big)$&
${}\frac{1}{7}$ &
${}0$           & 
                        &
                        &       
\\[1ex]\hline&&&&&&\\ [-2ex]
$\big(\rep{2n_0},\crep{n_3}\big)$,
$\big(\crep{n_1},\crep{n_2}\big)$&
&
${}0$           &
                        & 
                        &                       &
$\tN$
\\[1ex]\hline\hline 
\multicolumn{2}{|c||}{}&&&&&\\ [-2ex]
\multicolumn{2}{|l||}{spinorial repr.\hfill $\backslash$\hfill 
$\frac 14(49n_0+25 n_1+9 n_2+n_3)$} &
14            &
28           &
42           &
56           &
70
\\[1ex]\hline\hline \multicolumn{2}{|l||}{}&&&&&\\ [-2ex]
\multicolumn{2}{|l||}{$\big(\spin{\ga}{n_0}\big)$}&
${}\frac{4}{7}$&
${}\frac{3}{7}$&
${}\frac{2}{7}$&
${}\frac{1}{7}$&
${}0$  
\\[1ex]\hline\multicolumn{2}{|l||}{}&&&&&\\ [-2ex]
\multicolumn{2}{|l||}{$\big(\spin{\sm\ga}{n_0},\rep{n_1}\big)$}&
${}\frac{3}{7}$&
${}\frac{2}{7}$&
${}\frac{1}{7}$&
${}0$         &
\\[1ex]\hline\multicolumn{2}{|l||}{}&&&&&\\ [-2ex]
\multicolumn{2}{|l||}{$\big(\spin{\sm\ga}{n_0},\rep{n_2}\big)$,
$\big(\spin{\ga}{n_0},[\rep{n_1}]_2\big)$}&
${}\frac{2}{7}$& 
${}\frac{1}{7}$& 
${}0$         &
              &
\\[1ex]\hline\multicolumn{2}{|l||}{}&&&&&\\ [-2ex]
\multicolumn{2}{|l||}{$\big(\spin{\sm\ga}{n_0},\rep{n_3}\big)$,
$\big(\spin{\ga}{n_0},\rep{n_1},\rep{n_2}\big)$,
$\big(\spin{\sm\ga}{n_0},[\rep{n_1}]_3\big)$}
&
${}\frac{1}{7}$ &
${}0$           & 
                &
                &       
\\[1ex]\hline\multicolumn{2}{|l||}{}&&&&&\\ [-2ex]
\multicolumn{2}{|l||}{$\big(\spin{\sm\ga}{n_0},\crep{n_3}\big)$,
$\big(\spin{\ga}{n_0},\rep{n_1},\rep{n_3}\big)$,
$\big(\spin{\ga}{n_0},[\rep{n_2}]_2\big)$}
&
${}0$           &
                & 
                &
                &
$\tN$
\\[1ex]\hline
\end{tabular} 
\end{center} 
\captn{
For an even (odd) shift $v_{even}$ ($v_{odd}$) the first twisted
spectrum has even (odd) vectorial weight representation. The chirality
of the spinors in the spinorial weights is given by $\ga=(-)^{n_0}$
($\ga=(-)^{n_0+1})$ for an even (odd) shift. 
\labl{tb:1TwZ7spi}\labl{tb:1TwZ7vec}
}
\end{table}
}


\newcommand{\Zsix}{
\def\thetable{\ref{sc:ModZ6}.\arabic{table}} 
\setcounter{table}{0}
\begin{table} 
%
%
\[
\arry{| l | l |}{
\hline & \\[-2ex]  
\Intr_6\mbox{--I} & \gf = \frac 16\big(1,1, \mbox{-}2 \big)
\\[1ex] \hline\hline & \\[-2ex] 
\gf_{1,2} = \frac 16 & 
\big(\rep{2 n_0}, \rep{n_1}  \big) , 
\big(  \crep{n_2}, \rep{2n_3}\big) , 
\big( \crep{n_1}, \rep{n_2} \big)
\\[1ex] \hline & \\[-2ex] 
\gf_{3} = \mbox{-}\frac 13 & 
[\crep{n_1}]_2 , 
[\rep{n_2}]_2 ,
\big( \rep{n_1},  \rep{2n_3}\big) ,
\big(\rep{2 n_0},  \crep{n_2}  \big)
\\[1ex] \hline 
}
\]
\captn{
This table gives the untwisted spectrum of the $\Intr_6$ orbifold
models.  
\labl{tb:UnZ6}}
%
%
\begin{center} \begin{tabular}{| c || c | c | c | c | c |}
\hline &&&&&\\ [-2ex]
$\tN$ & $0$ & $\frac 16$ & $\frac 13$ & $\frac 12$ & $\frac 23$ 
\\[1ex]\hline\hline &&&&&\\ [-2ex]
states & $\ket{0}$ & $\ket{\undr{1}^k\undr{2}^{1\mbox{-}k}}$ & 
 $\ket{\undr{1}^k\undr{2}^{2\mbox{-}k}}, \ket{3}$ & 
$\ket{\undr{1}^k \undr{2}^{3\mbox{-}k}},  \ket{\undr{1}^k\undr{2}^{1\mbox{-}k}3}$ & 
$\ket{\undr{1}^k \undr{2}^{4\mbox{-}k}},  \ket{\undr{1}^k\undr{2}^{2\mbox{-}k}3}, 
\ket{\undr{3}},\ket{3^2}$  
\\[1ex]\hline&&&&&\\ [-2ex]
$\SU{2}$ hol. & $\rep{1}$ & $\crep{2}$ & $\crep{3}, \rep{1}$ & 
$\crep{4}, \crep{2}$ &  $\crep{5}, \crep{3}, \rep{1},\rep{1}$ 
\\[1ex]\hline&&&&&\\ [-2ex]
multi. & 1 & 2 & 4 & 6 & 10
\\[1ex]\hline
\end{tabular} \end{center}
\captn{
The possible values of $\tN$ for the first twisted sector of a
$\Intr_6$--I are $0, \frac 16, \ldots \frac 23$. The index structure of the
corresponding states, their $\SU{2}$ holonomy representations
and total multiplicities are indicated. 
\labl{tb:1stTwZ6multi}}
%
%
\begin{center}
\scalebox{1}{\mbox{
\begin{tabular}{| l || c | c | c | c | c |}
\hline &&&&&\\ [-2ex]
vectorial repr. \hfill $\backslash$ \hfill $n_1+4 n_2+9n_3$ &
\,6\,  & 18  &  30  & 42 & 54           
\\[1ex]\hline\hline &&&&&\\ [-2ex]
$\big(\spin{+}{n_3}\big)$  &
$\frac{2}{3}$&
$\frac{1}{2}$&
$\frac{1}{3}$&
$\frac{1}{6}$&
$0$  
\\[1ex]\hline&&&&&\\ [-2ex]
$\big( \crep{n_2}, \spin{-}{n_3} \big)$  &
$\frac{1}{2}$&
$\frac{1}{3}$&
$\frac{1}{6}$&
$0$           &
\\[1ex]\hline&&&&&\\ [-2ex]
$\big( \crep{n_1},\spin{-}{n_3} \big)$,
$\big( [\crep{n_2}]_{2},\spin{+}{n_3}\big)$ &
$\frac{1}{3}$&
$\frac{1}{6}$& 
$0$           &&       
\\[1ex]\hline&&&&&\\ [-2ex]
$\big( \rep{2n_0},\spin{-}{n_3}\big)$,
$\big( \crep{n_1},\crep{n_2},\spin{+}{n_3}\big)$, 
$\big( [\crep{n_2}]_{3},\spin{-}{n_3}\big)$
&
$\frac{1}{6}$&
$0$     &&&
\\[1ex]\hline&&&&&\\ [-2ex]
$\big( [\crep{n_1}]_{2},\spin{+}{n_3}\big)$,
$\big( \rep{2n_0},\crep{n_2},\spin{+}{n_3}\big)$, 
$\big( \rep{n_1},\spin{-}{n_3}\big)$ 
& 
$0$     &&&& $\tN$ 
\\[1ex]\hline  
\hline &&&&&\\ [-2ex]
spinorial repr. \hfill $\backslash$ \hfill 
$n_2+4 n_1+9n_0$ &
\,6\, & 18 & 30 & 42 & 54               
\\[1ex]\hline\hline &&&&&\\ [-2ex]
$\big( \spin{\ga}{n_0}\big)$  &
$\frac{2}{3}$&
$\frac{1}{2}$&
$\frac{1}{3}$&
$\frac{1}{6}$&
$0$
\\[1ex]\hline&&&&&\\ [-2ex]
$\big(\spin{\sm\ga}{n_0},\rep{n_1}\big)$  &
$\frac{1}{2}$&
$\frac{1}{3}$&
$\frac{1}{6}$&
$0$  &
\\[1ex]\hline&&&&&\\ [-2ex]
$\big(\spin{\sm\ga}{n_0},\rep{n_2}\big)$,
$\big(\spin{\ga}{n_0},[\rep{n_1}]_{2} \big)$&
$\frac{1}{3}$&
$\frac{1}{6}$& 
$0$  &&
\\[1ex]\hline&&&&&\\ [-2ex]
$\big(\spin{\sm\ga}{n_0},\rep{2 n_3}\big)$,
$\big(\spin{\ga}{n_0},\rep{n_1},\rep{n_2}\big)$,
$\big(\spin{\sm\ga}{n_0},[\rep{n_1}]_{3}\big)$
&
$\frac{1}{6}$&
$0$   &&& 
\\[1ex]\hline&&&&&\\ [-2ex]
$\big(\spin{\ga}{n_0}, [\rep{n_2}]_{2}\big)$,
$\big(\spin{\ga}{n_0},\rep{n_1},\rep{2n_3}\big)$,
$\big(\spin{\sm\ga}{n_0},\crep{n_2}\big)$
&
$0$ &&&& $\tN$ 
\\[1ex]\hline 
\end{tabular} 
}}
\end{center}
\captn{
The spectrum of the first twisted sector of $\Intr_6$--I orbifold
models can be read off from this table. To determine the appropriate
multiplicities it should be combined with table above. 
The model dependent chirality reads $\ga =(-)^{n_0}$. 
When $n_0$ or $n_3$ is zero, only those states should be 
kept with positive chirality.  
\labl{tb:1stTwZ6}
}
\end{table}
}


\newcommand{\Zeight}{
\def\thetable{\ref{sc:ModZ8}.\arabic{table}} 
\setcounter{table}{0}
\begin{table} 
%
%
\[
\arry{| l | l |}{
\hline & \\[-2ex]  
\Intr_8\mbox{--I} & \gf = \frac 18\big(1,2, \mbox{-}3 \big)
\\[1ex] \hline \hline  & \\[-2ex] 
\gf_{1} = \frac 18 & 
\big(\rep{2 n_0}, \rep{n_1}  \big) , 
\big(   \crep{n_3},\rep{2n_4}\big) , 
\big( \crep{n_1}, \rep{n_2}, \big) , 
\big(  \crep{n_2}, \rep{n_3}\big) 
\\[1ex] \hline & \\[-2ex] 
\gf_{2} = \frac 14 & 
\big( [\rep{n_1}]_2  \big) , 
\big(   [\crep{n_3}]_2\big) , 
\big(\rep{2n_0},  \rep{n_2} \big) , 
\big(  \crep{n_2}, \rep{2n_4}\big) 
, \big( \crep{n_1},  \rep{n_3}\big) 
\\[1ex] \hline & \\[-2ex] 
\gf_{3} = \mbox{-}\frac 38 & 
\big(\rep{2n_0},   \crep{n_3}\big) , 
\big( \rep{n_1},  \rep{2n_4}\big) , 
\big( \crep{n_1}, \crep{n_2} \big) , 
\big( \rep{n_2}, \rep{n_3} \big) 
\\[1ex] \hline 
}
\]
\captn{
This table gives the untwisted spectrum of the $\Intr_8$--I orbifold
models. 
\labl{tb:SpecUnZ8}}
%
%
\begin{center} 
\begin{tabular}{| c || c | c | c | c | c | c |}
\hline &&&&&&\\ [-2ex]
$\tN$ & $0$ & $\frac 18$ & $\frac 14$ & $\frac 38$ & $\frac 12$ & $\frac 58$
\\[1ex]\hline\hline &&&&&&\\ [-2ex]
states          & 
$\ket{0}$       & 
$\ket{\undr{1}}$ & 
$\ket{\undr{1}^2}, \ket{\undr{2}}$ & 
$\ket{\undr{1}^3}, \ket{\undr{1}\undr{2}}, \ket{3}$ & 
$\ket{\undr{1}^4}, \ket{\undr{1}^2\undr{2}}, \ket{\undr{2}^2},
\ket{\undr{1}3}$&
$\ket{\undr{1}^5}, \ket{\undr{1}^3\undr{2}}, \ket{\undr{1}\undr{2}^2},
\ket{\undr{1}^23}, \ket{\undr{2}3},\ket{\undr{3}}$
\\[1ex]\hline&&&&&&\\ [-2ex]
multi. & 1 & 1 & 2 & 3 & 4 & 6
\\[1ex]\hline
\end{tabular} 
\end{center}
\captn{
The possible values of $\tN$ for the first twisted sector of a
$\Intr_8$--I are $0, \frac 18, \ldots \frac 58$. The index structure of the
corresponding states and total multiplicities are indicated. The holonomy
group is trivially $U(1)^3$.
\labl{tb:1stTwZ8Imulti}}
%
%
\begin{center}
\begin{tabular}{| l || c | c | c | c | c | c |}
\hline &&&&&&\\ [-2ex]
vectorial repr. \hfill $\backslash$ \hfill $n_1+4 n_2+9n_3+16n_4$ &
14           &
30           &
46           &
62           &
78           &
94      
\\[1ex]\hline\hline &&&&&&\\ [-2ex]
$\big(\spin{+}{n_4}\big)$
&
${}\frac{5}{8}$&
${}\frac{4}{8}$&
${}\frac{3}{8}$&
${}\frac{2}{8}$&
${}\frac{1}{8}$&
${}0$  
\\[1ex]\hline&&&&&&\\ [-2ex]
$\big(\crep{n_3},\spin{-}{n_4}\big)$&
${}\frac{4}{8}$&
${}\frac{3}{8}$&
${}\frac{2}{8}$&
${}\frac{1}{8}$&
${}0$         &
\\[1ex]\hline&&&&&&\\ [-2ex]
$\big(\crep{n_2},\spin{-}{n_4}\big)$,
$\big([\crep{n_3}]_{2},\spin{+}{n_4}\big)$&
${}\frac{3}{8}$&
${}\frac{2}{8}$& 
${}\frac{1}{8}$& 
${}0$         &
                      &        
\\[1ex]\hline&&&&&&\\ [-2ex]
$\big(\crep{n_1},\spin{-}{n_4}\big)$,
$\big(\crep{n_2},\crep{n_3},\spin{+}{n_4}\big)$,
$\big([\crep{n_3}]_{3},\spin{-}{n_4}\big)$&
${}\frac{2}{8}$&
${}\frac{1}{8}$&
${}0$   &
                &
                &              
\\[1ex]\hline&&&&&&\\ [-2ex]
$\big(\rep{2n_0},\spin{-}{n_4}\big)$,
$\big(\crep{n_1},\crep{n_3},\spin{+}{n_4}\big)$,
$\big([\crep{n_2}]_{2},\spin{+}{n_4}\big)$,
$\big(\crep{n_2},[\crep{n_3}]_{2},\spin{-}{n_4}\big)$
&
${}\frac{1}{8}$ &
${}0$           &
                        & 
                        &
                        &       
\\[1ex]\hline&&&&&&\\ [-2ex]
$\big(\rep{2n_0},\crep{n_3},\spin{+}{n_4}\big)$,
$\big(\crep{n_1},\crep{n_2},\spin{+}{n_4}\big)$,
$\big([\crep{n_2}]_{2},\crep{n_3},\spin{-}{n_4}\big)$,
$\big(\rep{n_1},\spin{-}{n_4}\big)$ &
${}0$   &
        &
        & 
        &
        &
$\tilde N$
\\[1ex]\hline
\hline &&&&&&\\ [-2ex]
spinorial repr. \hfill $\backslash$ \hfill $n_3+4 n_2+9n_1+16n_0$ &
14            &
30           &
46           &
62           &
78           &
94      
\\[1ex]\hline\hline &&&&&&\\ [-2ex]
$\big(\spin{\ga}{n_0} \big)$&
${}\frac{5}{8}$&
${}\frac{4}{8}$&
${}\frac{3}{8}$&
${}\frac{2}{8}$&
${}\frac{1}{8}$&
${}0$  
\\[1ex]\hline&&&&&&\\ [-2ex]
$\big(\spin{\sm\ga}{n_0},\rep{n_1} \big)$&
${}\frac{4}{8}$&
${}\frac{3}{8}$&
${}\frac{2}{8}$&
${}\frac{1}{8}$&
${}0$         &
\\[1ex]\hline&&&&&&\\ [-2ex]
$\big(\spin{\sm\ga}{n_0},\rep{n_2} \big)$,
$\big(\spin{\ga}{n_0},[\rep{n_1}]_{2} \big)$&
${}\frac{3}{8}$&
${}\frac{2}{8}$& 
${}\frac{1}{8}$& 
${}0$         &
                      &        
\\[1ex]\hline&&&&&&\\ [-2ex]
$\big(\spin{\sm\ga}{n_0},\rep{n_3} \big)$,
$\big(\spin{\ga}{n_0},\rep{n_1},\rep{n_2}\big)$,
$\big(\spin{\sm\ga}{n_0},[\rep{n_1}]_{3} \big)$&
${}\frac{2}{8}$&
${}\frac{1}{8}$&
${}0$   &
                &
                &              
\\[1ex]\hline&&&&&&\\ [-2ex]
$\big(\spin{\sm\ga}{n_0},\rep{2n_4} \big)$,
$\big(\spin{\ga}{n_0},\rep{n_1},\rep{n_3}\big)$,
$\big(\spin{\ga}{n_0},[\rep{n_2}]_{2} \big)$,
$\big(\spin{\sm\ga}{n_0},[\rep{n_1}]_{2},\rep{n_2} \big)$&
${}\frac{1}{8}$ &
${}0$           &
                        & 
                        &
                        &
\\[1ex]\hline&&&&&&\\ [-2ex]
$\big(\spin{\ga}{n_0},\rep{n_1},\rep{2n_4} \big)$,
$\big(\spin{\ga}{n_0},\rep{n_2},\rep{n_3}\big)$,
$\big(\spin{\ga}{n_0},\rep{n_1},[\rep{n_2}]_{2} \big)$,
$\big(\spin{\sm\ga}{n_0},\crep{n_3} \big)$&
${}0$   &
                &
                & 
                &
                &
$\tilde N$                   
\\[1ex]\hline
\end{tabular} 
\end{center}
\captn{\labl{tb:1stTwZ8I}
This table gives the first twisted sector 
states for the $\Intr_8$--I model. $\ga=(-1)^{n_0}$.}
\end{table}
}

\section{$\boldsymbol{\Intr_7}$ models}
\labl{sc:ModZ7}\setcounter{equation}{0}

This appendix is devoted the $\Intr_7$ models with spacetime
shift $\gf = \frac 17(1,2,\mbox{-}3)$. The tables
\ref{tb:SpecUnZ7}--\ref{tb:1TwZ7vec} on the next page 
give their complete spectrum.  By Weyl reflections and additions of
roots we can bring the gauge shift $v$ in  the standard forms: 
\equ{
v_{even} = \frac 17 \big( 0^{n_0}, 1^{n_1}, 2^{n_2}, 3^{n_3} \big),
\qquad 
v_{odd} = \frac 17 \big( 0^{n_0}, 1^{n_1}, 2^{n_2}, 3^{n_3-1}, \mbox{-}4 \big).
\labl{Z7shifts}
}
Notice that $v_{odd} = v_{even} - d_1$ with $d_1 = (0^{15}, 1)$. 
These shifts are not equivalent since for the modular invariance
condition  $3/7$ and $\sm 4/7$ are not equivalent: For the shifts
$v_{even}$ the modular invariance condition is satisfied by
null--solutions, that have the form 
\equ{
\gn = (14 p_1, 7p_2, 14 p_3) + q_1(4, \mbox{-}1,0) 
+ q_2 (9, 0, \mbox{-}1), 
\labl{Z7null} 
}
while the shifts $v_{odd}$ are obtained by adding the shift $(3,1,0)$ 
to the null--solutions. Both shifts gives rise to the gauge group: 
\equ{
\SO{2n_0} \times \U{n_1} \times \U{n_2} \times \U{n_3}.
}
In table \ref{tb:SpecUnZ7} we have given the untwisted matter
spectrum.

The twisted matter is given by three irreducible sectors: The first, second
and fourth twisted sector all carry the same four dimensional
chirality (\ref{chirality}). The vectorial weights do not contain
spinor representations and therefore the GSO condition acts as a 
projection. For these representations in the first twisted sector this
projection depends on which shift is used: 
\equ{
\text{First twisted}~:~~ 
v_{even} ~:~ \frac 12 \sum_{a=0}^3 k_a \equiv 0, 
\qquad 
v_{odd} ~:~ \frac 12 \sum_{a=0}^3 k_a \equiv \frac 12. 
}
We call the corresponding vectorial weights even and odd,
respectively. The spinorial weights always contain a spinor
representation and hence the GSO selects as usual only the chirality
of this spinor: For even shifts we have $\ga_{even} = (-)^{n_0}$,
while for odd $\ga_{odd} = (-)^{n_0+1}$. The content of the first
twisted sector is summarized in tables \ref{tb:1stTwZ7Imulti} 
and \ref{tb:1TwZ7spi}.

The second twisted sector content is obtained simply by these
tables by employing the transformation matrix 
\equ{
S_2 = \pmtrx{
\arry{c|c|c|c}{
\Id_{n_0} &&& \\\hline && \Id_{n_1} & \\\hline 
&&& \sm \Id_{n_2} \\\hline & \sm \Id_{n_3} &&
}}, 
\labl{SpZ7}
}
and results in the replacements: 
$\rep{n_1} \ra \crep{n_3} \ra \rep{n_2} \ra \rep{n_1}$. 
However, the GSO condition for the vectorial weights is different: The
vectorial second twisted sector is even (odd) if $n_2+n_3$ is even
(odd). For the spinorial weights the chirality are selected by 
$\ga = (-)^{n_0}$ for both even and odd shifts. The bosonic
excitations are interchanged in table \ref{tb:1stTwZ7Imulti} 
as: $1 \ra 3 \ra 2 \ra 1$. The transformation matrix  
\equ{
S_4 = (S_2)^2 = \pmtrx{
\arry{c|c|c|c}{
\Id_{n_0} &&& \\\hline &&& \sm\Id_{n_1}  \\\hline 
&  \Id_{n_2} && \\\hline && \sm \Id_{n_3} &
}}.
}
indicates that the fourth twisted sector is obtained from the first
twisted sector by $\rep{n_1} \ra \rep{n_2} \ra \crep{n_3} \ra \rep{n_1}$.  
In this case the even (odd) vectorial weights are
required for even (odd) $n_1+n_2$. The chirality of the spinorial
weights reads  $\alpha=(-)^{n_2+n_2+n_3}$. The interchange of the
bosonic excitations in table  \ref{tb:1stTwZ7Imulti}  reads: 
$1 \ra 2 \ra 3 \ra 1$.

In subsection \ref{sc:hettypI} we consider the type--I/heterotic
duality. To facilitate that discussion we give the irreducible
anomalies of the untwisted states, which are listed in table
\ref{tb:SpecUnZ7}. The corresponding anomaly polynomial reads 
\equ{\arry{rcl}{
I_{6|u\ irr} &= & \dsp  
- \frac 16 \big( 4 - 2n_0 - n_1 + 2n_2 \big) \tr\, F_1^3 
- \frac 16 \big( 4 - 2n_0 + n_1 -2n_2 + 2n_3  \big) \tr\, F_2^3 
\\[2ex] & & \dsp 
- \frac 16 \big( -4 + 2n_0 -2n_1 + n_3  \big) \tr\, F_3^3 
}
\labl{IrrAnomZ7} 
} 
where $F_a$ are the $\U{n_a}$ field strengths. It is not difficult to
show that with the constraint $n_0+n_1+n_2+n_3=16$, there is only a
unique solution for which these irreducible anomalies are absent: 
$n_0=n_1=n_2=n_3=4$.


\Zseven 

\newpage

\section{$\boldsymbol{\Intr_{6}}$ models}
\labl{sc:ModZ6}
\setcounter{equation}{0}

The $\Intr_6\mbox{--I}$ models have the space shift 
$\gf = \frac 16\big(1,1, \mbox{-}2 \big)$. Their spectra are
listed in tables \ref{tb:UnZ6}--\ref{tb:1stTwZ6} on the next page.  
The gauge shift and the level matching condition read for the
$\Intr_6$ theory 
\equ{
v = \frac 16 \big( 0^{n_0}, 1^{n_1}, 2^{n_2}, 3^{n_{3}} \big), 
\qquad 3 v^2 =  
\frac 1{12} n_1 + \frac 13 n_2 + \frac 34 n_3 \equiv \frac 12. 
\labl{LevelMatch6}
}
The  four dimensional gauge group becomes 
\equ{
\SO{2n_0} \times \U{n_1} \times \U{n_2} \times \SO{2 n_3}.
}
The null--solutions of the level matching condition are given by 
\equ{
3 v_\gn^2 \equiv 0 \quad \Leftrightarrow \quad 
\gn = (12 p_1, 3p_2, 4 p_3) + q_1(4, \mbox{-}1, 0) + 
q_2( 9, 0, -1). 
}
There are two inequivalent way a $\Intr_6$ acts on the
six--torus distinguished by the form of the spacetime shift vector: 
A particular solution of the level matching condition 
\eqref{LevelMatch6} is given by $ n = (6, 0, 0)$.

The matter spectrum of $\Intr_6$--I models built  as follows: 
The untwisted matter is given in table \ref{tb:UnZ6}. There are five
twisted sector of which the first is conjugate to the fifth, and the
second conjugate to the fourth. The third twisted sector obtained from
the six dimensional $\Intr_2$ sector described in table
\ref{tb:SpecTwT4Z2} in subsection \ref{sc:6Dmatter}. The first twisted
sector has been collected in tables \ref{tb:1stTwZ6multi} and
\ref{tb:1stTwZ6}. Finally the second twisted sector arises as the first
twisted sector of a $\Intr_3$ orbifold with gauge shift
\equ{
\tv_2^{vec} = \frac 13 \big( 0^{n_0+n_3}, 1^{n_1+n_2} \big), 
}
in the standard $\Intr_3$ ordering. The $\Intr_3$ twisted matter has
been collected in table \ref{tb:SpecT6Z3} (or \ref{tb:TwStates}). 
To interpolate between the $\Intr_6$ to the $\Intr_3$ ordering of
shift vectors, the matrix  
\equ{
S_2 = \pmtrx{
\arry{c|c|c|c}{
\Id_{n_0} &&& \\\hline && \Id_{n_1} & \\\hline 
&&& \sm \Id_{n_2} \\\hline & \Id_{n_3} &&
}}
}
has been employed. Both the second and third twisted sectors are
subject to appropriate identifications and further orbifolding or
projections, depending to which fixed points these states are
associated.


\Zsix 

\newpage

\section{$\boldsymbol{\Intr_{8}}$ models}
\labl{sc:ModZ8}
\setcounter{equation}{0}

The $\Intr_8\mbox{--I}$ models have space shift 
$\gf = \frac 18\big(1,2, \mbox{-}3 \big)$. Their spectra are listed in
tables \ref{tb:SpecUnZ8}--\ref{tb:1stTwZ8I} on the next page. 
The gauge shift and the level matching condition
read for the $\Intr_8$ theory 
\equ{
v = \frac 16 \big( 0^{n_0}, 1^{n_1}, 2^{n_2}, 3^{n_{3}}, 4^{n_4} \big), 
\qquad 
4 v^2 = 
\frac 1{16} n_1 + \frac 14 n_2 + \frac 9{16}n_3+ n_4 \equiv \frac 78. 
\labl{LevelMatch8}
}
The resulting gauge group becomes 
\equ{
\SO{2n_0} \times \U{n_1}\times  \U{n_2} \times 
\U{n_3} \times \SO{2 n_4}.
}
The null--solutions of the level matching condition are given by 
\equ{
4 v_\gn^2 \equiv 0 \quad \Leftrightarrow \quad 
\gn = (16 p_1, 4p_2, 16 p_3, p_4) + q_1(4, \mbox{-}1, 0, 0) + 
q_2( 9, 0, \mbox{-}1,0). 
}
A particular solution to the modular invariance condition 
\eqref{LevelMatch8} is given by $n = (5, 0, 1, 0)$.

The untwisted sector is listed in table \ref{tb:SpecUnZ8}. 
The twisted sector that describe four dimensional matter with the same
chirality as the untwisted states are the first, second and fifth
twisted sectors. The first twisted sector states are given by tables
\ref{tb:1stTwZ8Imulti} and \ref{tb:1stTwZ8I}. The fifth twisted sector
is related to the first twisted sector by the matrix 
\equ{
S_5 = \pmtrx{
\arry{ c|c|c|c|c}{
\Id_{n_0} & & & & \\\hline 
& & & \sm \Id_{n_3} & \\ \hline 
& & \Id_{n_2} & & \\ \hline 
& \sm \Id_{n_1} & & & \\ \hline 
& & & & \Id_{n_4} 
}
}
}
via $\tv^{vec}_5 S_5 = v$, and in addition the spacetime indices need
to be interchanged. Concretely, this means that in table
\ref{tb:1stTwZ8Imulti} we interchange $1 \leftrightarrow 3$, and in table 
\ref{tb:1stTwZ8I} we map: $\rep{n_1} \ra \crep{n_3}$ and 
$\rep{n_3} \ra \crep{n_1}$. 
The chirality of the spinorial representations are also modified
due to the presence of a non--trivial $d_5$. In particular
the $+$ chirality of the spinorial states related to vectorial weights 
in table \ref{tb:1stTwZ8I} is replaced by $(-1)^{n_1+n_2}$, while the 
$\alpha$ chirality appearing in the spinorial states related to spinorial
weights is replaced by $\alpha=(-1)^{n_0+n_2+n_3}$.

The second twisted sector is not irreducible as it is obtained from the
first twisted sector of $\Intr_4$ orbifolds. The relevant information
for these spectra are given in 
tables \ref{tb:1stTwT6Z4multi} and \ref{tb:1stTwT6Z4}. However, to use
these tables for the second twisted sector of the $\Intr_8$--I model
one should interchange the space indices $2 \ra 3$ in table
\ref{tb:1stTwT6Z4multi} and use the matrix 
\equ{
S_2 = \pmtrx{
\arry{c|c|c|c|c}{
\Id_{n_0} & & & & \\ \hline 
& & \Id_{n_2} & & \\ \hline 
& & & & \Id_{n_4} \\ \hline 
& & & \sm \Id_{n_3} & \\ \hline 
& \Id_{n_2} & & &
}
}
}
to bring $\tv^{vec}_2$ in the standard form of a $\Intr_4$ shift.

\Zeight


\end{document}